\documentclass[fleqn,usenatbib]{mnras}

\usepackage{newtxtext,newtxmath}

\usepackage[T1]{fontenc}

\DeclareRobustCommand{\VAN}[3]{#2}
\let\VANthebibliography\thebibliography
\def\thebibliography{\DeclareRobustCommand{\VAN}[3]{##3}\VANthebibliography}

\usepackage{graphicx}	
\usepackage{amsmath}	

\newcommand{\nudot}[0]{$\dot\nu$}

\title[Measuring Profile Variations with a 2-D GP]{Measuring pulsar profile variations with 2-D Gaussian process regression}

\author[M. J. Keith et al.]{
Michael J. Keith,$^{1}$\thanks{E-mail: mkeith@pulsarastronomy.net}
Ren\'ee Spiewak,$^{1}$
Andrew G. Lyne,$^{1}$
Patrick Weltevrede,$^{1}$
Danai Antonopoulou,$^{1}$\newauthor
and Ben Stappers$^{1}$
\\
$^1$Jodrell Bank Centre for Astrophysics, Department of Physics and Astronomy, The University of Manchester, Manchester M13 9PL, UK
}

\date{Accepted XXX. Received YYY; in original form ZZZ}

\pubyear{2024}

\begin{document}
\label{firstpage}
\pagerange{\pageref{firstpage}--\pageref{lastpage}}
\maketitle

\begin{abstract}
Time-correlated variations in the pulse profiles of radio pulsars provide insights into changes in their magnetospheres. For a small number of pulsars ($\sim20$), these variations have been shown to correlate with spin-down rate. Many of these profile changes involve small (few percent) variations in the relative intensity of different profile components, and hence tools such as Gaussian process regression have been employed to separate the time-correlated profile variation from intrinsic noise.
In this paper, we present a computationally efficient approximation of a 2-D Gaussian process model that enhances sensitivity by simultaneously tracking time- and phase-correlated signals. 
Applying this model to 26 pulsars observed at the Jodrell Bank Observatory, we detect significant profile shape variations in 21 pulsars. Using principal component analysis, we confirm spin-down correlated shape variations in 11 pulsars where this had been previously observed. Additionally, we find evidence of spin-down correlated shape changes in 7 pulsars for the first time (PSRs B0105+65, B0611+22, B0626+24, B1740$-$03, B1826$-$17, B1917+00, and B2148+63).
We look in greater detail at PSR B0740$-$28, where the correlation between profile shape and spin-down itself seems to switch between quasi-stable states. Notably the profile shape associated with greater spin-down seems to invert at times, presenting a challenge to our understanding of the physical processes at work.

\end{abstract}

\begin{keywords}
pulsars: general -- methods: data analysis -- pulsars: individual: B0740$-$28
\end{keywords}



\section{Introduction}

Neutron stars host very strong magnetic fields, possibly even reaching up to $10^{16}$ G in the interior of highly-magnetised ones (i.e. magnetars), and are thus surrounded by an extended magnetosphere. The structure of the magnetosphere dictates plasma production and acceleration, and in turn most of the observed electromagnetic emission that allows us to study these compact objects. For isolated neutron stars, the magnetospheric energy losses are considered the dominant mechanism by which the star's rotational energy decreases, leading to an observed spin-down. 

The majority of known neutron stars are detected as radio pulsars, with the beamed radio emission detectable on Earth for only part of the pulsar's rotational phase. By creating a time-averaged, phase-binned histogram of the pulses intensities, a pulse profile can be constructed for each source.
For most pulsars, whilst individual pulses have unpredictable shapes that do not reflect the long-term average, the profile converges towards a stable shape after averaging a few hundred rotations, with each pulsar associated with it's own unique pulse shape \citep{helfand75}.
However, not all pulsars have a single stable profile, and the average of a sequence of pulses may converge towards two or more stable states \citep{backer70a}, often called `modes'.
The timescales associated with this moding phenomenon extend over the full range of our sensitivity to such things, i.e. from a few pulses (below which we cannot reliably identify states) up to years, potentially exceeding the length of time we have observed a particular pulsar.

\citet[hereafter L10]{L10} identified changes in the observed pulse profile of 6 pulsars which showed a clear correlation with the pulsar spin-down rate (\nudot), often in the form of quasi-periodic switching of states.
The pulse profile changes were characterised by a specific shape parameter for each pulsar, typically some measure of the width of the pulse.
The shape parameter is averaged over partially overlapping time windows of 100-400 days so that short term fluctuations are smoothed out.

There are broadly two approaches that may be taken when trying to interpret the observed pulse profile changes.
One approach is to make use of pulsar emission models, and directly interpret the data in the light of these models.
For instance, one can try to separate the components of the pulse profile by their classification in canonical model of central core emission surrounded by one or two nested cones \citep{rankin83}, which has been highly successful in explaining a range of pulsar emission phenomenology \citep[e.g.][]{mitra11,Olszanski19,corecone_edot}.
The other approach is to perform a model-independent analysis of the observations, free from assumptions about the emission, and later consider the application to the models of pulsars.
Although both approaches are beneficial, in this paper we will focus on a new method for the model-independent study of the pulse shape variations.

\citet{aris11} first developed a model-independent method for the detection of underlying pulse shape variations by making use of Gaussian Process  (GP) regression.
\citet{brook16} used a further refined version of this algorithm to discover time-correlated pulse shape variations in 5 southern pulsars, one of which showed clear correlations with \nudot.
In brief the algorithm is as follows.
The pulse profiles are aligned in phase and normalised, before the average profile is subtracted from each profile.
This results in a `difference map', which is often dominated by white noise, both due to the instrument and intrinsic to the pulsar due to the finite number of pulses that have been averaged in each observation.
In order to separate the noise from the underlying process, each pulse phase bin is treated as a GP, composed of a time-correlated term and a white noise term.
Reconstructing the underlying GP from the model produces a noise-free smooth model of the pulse shape variations.
\citet[hereafter S22]{S22} applied this same technique to the 17 pulsars studied in L10, and showed variations correlated with $\dot{\nu}$ in one additional pulsar.

The approach of \citet{brook16} and S22, is to align the profiles and fit the Gaussian process to each phase bin independently. This works well, but has two potential downsides.
Firstly, pulse profile variations are typically of the form of a varying profile component wider than the phase resolution of the observation, and so the changes show correlations in pulse phase.
The assumption of independent phase bins means that these phase correlations are ignored, hence the more phase bins are used, the less well modelled the changes in time may appear. 
Secondly, the GP kernel hyperparameters, i.e. the length scale of the correlations, varies from phase bin to phase bin, which can make it more difficult to interpret the results. For example, where there is little signal the length scale often tends towards the upper bound and `smooth' away the noise in the reconstruction.
This can lead to the impression of a lower noise level than is really present when comparing to phase bins that have a shorter timescale.
It can also give the impression that the variations are constrained to a narrower range of phase bins than may be in reality.
Alternatively if length scale tends towards short values in some phase bins, the GP mean will show large fluctuations that detract from the appearance of the real smooth variation in other phase bins of pulsar. 

A potential improvement to this may be to use a 2-D GP model in both the time and phase direction.
Indeed, \citet{jk19} have shown that it is possible to model pulse profiles in the phase dimension using GP regression, even though pulse profiles are not strictly GPs.
Combining the two 1-D models of \citet{jk19} and \citet{brook16} into a 2-D model is not conceptually difficult, rather the limiting factor is the computational cost of traditional 2-D GP regression, which is prohibitive for all but the smallest datasets.

Recently the \textsc{Celerite} algorithm for fast 1-D GP regression has been demonstrated to be much more computationally efficient compared to traditional GP algorithms \citep{celerite}.
In particular, the computational complexity of \textsc{Celerite} scales linearly with the size of the dataset, compared to cubic scaling with more traditional methods.
These benefits come at the expense of requiring that the kernel of the covariance function in the GP is constructed in the form of a particular sum of complex exponential terms.
Whilst seemingly restrictive, this actually allows for a wide range of kernels that are applicable to many real-world problems in astrophysics \citep{celerite}.
In this paper we will demonstrate how the 1-D \textsc{Celerite} kernel can be used to synthesise a 2-D GP model for pulsar profile variations.

\section{Methodology}
\subsection{1-d interpretation of the pulse profile stack}
\label{sec:2d21d}
A sequence of pulse profiles from separate observations is usually considered as a 2-D `stack' of pulses, with rotational phase as one dimension and observation epoch as the second dimension.
However, it is also natural to `flatten' the pulse profile stack by considering them to be a train of pulses.
In fact, if we scale by the period of the pulsar, $P$, the phase axis of our profiles can be treated as a time dimension, and hence by shifting each profile to the time of the observation we can turn our pulse profile stack into a non-continuous 1-D time-series.
Note that this is not exactly equivalent to `unfolding' the observed data into the original pulsar lightcurve, since each pulse in this time-series is still the average pulse profile of a full observation.
For a pulsar observed at times $T$ we could write a new time axis
\[
t(\phi,T) = \phi P + T,
\]
where $\phi$ is the pulse phase.
The covariance function of such a dataset will have a combination of a narrow correlation between phase bins, plus a periodic term (with period $P$) as data taken at the same phase in later pulses is highly correlated due to the consistent pulse shape.

\begin{figure}
    \centering
    \includegraphics[width=8.5cm]{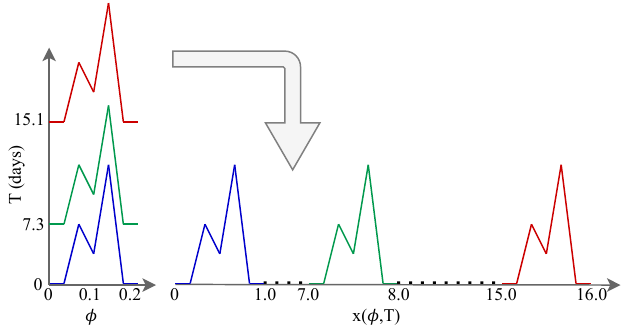}
    \caption{Left: The arrangement of three pulse profiles in a `pulse stack' in terms of phase, $\phi$, and time, $T$. Right: The arrangement of pulse profile data after converting to our new combined coordinate, $x(\phi,T)$ (See Equation \ref{eq1} for details.}.
    \label{fig:transform}
\end{figure}

However, this choice of our time-axis leads to extremely large values for $t$ with most of the data separated by very large gaps, which can lead to numerical problems due to the difference in scale  between the variations in phase and between observations (up to 10 orders of magnitude between e.g. microseconds and years).
Therefore we choose to artificially compress the time axis by scaling both $T$ and $\phi P$, whilst being careful not to break the periodic nature of the dataset.
Hence, we define $T$ to be the nearest integer number of days between the start of an observation and the first observation in the dataset, and effectively set the period of the pulsar to unity.
This requires that there is no more than one observation per day, but in practice observations are typically separated by tens of days and we can average closer observations if needed.
We also choose to discard data points outside of the phase region of interest, which we term the `on-pulse window'. 
Data outside of the on-pulse window do not help define the pulse shape, and actually may detract from the GP model of the phase dependence since the data are not really a stationary GP, for example,  the properties of the light-curve change significantly between the on and off-pulse regions.
Specifically, we define the $x$ values for the input to our GP by
\begin{equation}
    x(\phi,T) = \frac{\phi}{\delta \phi_\mathrm{onpulse}} + \left\lfloor \frac{T}{\mathrm{days}}\right\rfloor, \label{eq1}
\end{equation}
where $\delta \phi_\mathrm{onpulse}$ is the phase width of the on-pulse window.
This has the effect of making each on-pulse window have a unit width, and each day between observations also unit width, keeping the overall dimensions small.
A schematic representation of this transformation is shown in Figure \ref{fig:transform}.
Although it distorts our interpretation of pulse phase, it will turn out to be computationally advantageous to keep the width of the on-pulse window close to the repeat period of the observations.

\subsection{The Kernel}
The approach of Section \ref{sec:2d21d} for converting the 2-D dataset to a periodic 1-D dataset allows us to use fast 1-D GP solutions, however it does not provide the full functionality of a 2-D model.
The primary restriction is that the kernel cannot be an arbitrary 2-D shape, but must be the outer product of two 1-D kernels.
In practice we find this is sufficient for the application of measuring pulsar shape variations.

The GP kernel defines the shape of the covariance function of the data for a given lag, $\delta x$.
We want to construct a kernel that is composed of the product of a phase kernel $\Theta(\delta x)$, that represents the correlation in $\phi$ within a profile and is periodic on a period defined by angular frequency $\omega_0$, and a time kernel $\tau(\delta x)$ that represents the correlation over different observation epochs.  For our case we have chosen $x$ so that the periodicity is unity, and hence $\omega_0=2\pi$, but we first derive the kernel for general $\omega_0$.
The complete covariance kernel is the product
\[
k(\delta x) = \Theta(\delta x)\tau(\delta x).
\]
An example of such a kernel can be seen in the top panel of Figure~\ref{KernelExample}. 
\textsc{Celerite} kernels are of the form
\begin{align}
\begin{split}
K(\delta x) &= \sum_{n=1}^N \frac{1}{2}\left[
    (a_n + i\,b_n)\,e^{-(c_n+i\,d_n)\,\delta x} \right.\\
    &~~~~~~~~~~+\left.(a_n - i\,b_n)\,e^{-(c_n-i\,d_n)\,\delta x}\label{celerite_kernel}
\right],
\end{split}
\end{align}
where $a_n$, $b_n$, $c_n$, and $d_n$ are the coefficients of the kernel derived from the model hyperparameters.
This form can be used to implement a periodic kernel constructed as a Fourier series.
Therefore we define $\Theta(\delta x)$ in terms of its Fourier coefficients,
\[
\theta_n = \int\limits_{-\infty}^{\infty} \Theta(x) e^{-in\omega_0 x} \mathrm{d}x.
\]
Now, provided that $\Theta(\delta x)$ is both symmetric and real valued, we can approximate any kernel with a Fourier series of $N$ coefficients,
\begin{equation}
    k(\delta x) = \tau(\delta x) \left[\theta_0 + \sum_{n=1}^N
    \theta_n \, (e^{-i\,\omega_0 n\,\delta x} +e^{+i\,\omega_0 n\,\delta x}) \right].
\end{equation}
This can be implemented as a \textsc{Celerite} kernel provided we can also implement $\tau(\delta x)$ as a \textsc{Celerite} kernel.
In general $\tau(\delta x)$ has multiple terms and so to simplify the algebra we can consider only a single term in $\tau(\delta x)$, though we can simply sum over multiple terms if required.
In practice it quickly becomes computationally expensive to use $\tau(\delta x)$ with more than one \textsc{Celerite} term, but we have found that a single term $\tau(\delta x)$ is sufficient.
For this discussion we assume that $\tau(\delta x)$ can be written as
\[
\tau(\delta x) = \frac{1}{2}\left[
    (p + i\,q)\,e^{-(r+i\,s)\,\delta x} +
    (p - i\,q)\,e^{-(r-i\,s)\,\delta x}
    \right],
\]
with $p$,$q$,$r$, and $s$ the \textsc{Celerite} kernel parameters for $\tau(\delta x)$.
Now our complete kernel becomes
\begin{align}
\begin{split}
k(\delta x) &=  \frac{1}{2}\left[\sum_{n=1}^N
     \theta_n \, (p + i\,q)e^{-(r+i\,(\omega_0 n-s))\,\delta x} \right.\\
     &~~~~\left. +\, \theta_n(p - i\,q)e^{-(r-i\,(\omega_0 n-s))\,\delta x}\right]\\
    &+\frac{1}{2}\left[\sum_{n=1}^N \theta_n \, (p + i\,q)e^{-(r+i\,(\omega_0 n+s))\,\delta x} \right.\\
    &~~~~\left. + \,\theta_n(p - i\,q)e^{-(r-i\,(\omega_0 n+s))\,\delta x}\right]\\
    & +  \frac{1}{2}\left[
    \theta_0(p + i\,q)\,e^{-(r+i\,s)\,\delta x} +
    \theta_0(p - i\,q)\,e^{-(r-i\,s)\,\delta x}
    \right]
    \end{split}
\end{align}
By comparison with Equation \ref{celerite_kernel} it can be seen that this can be implemented with $2N+1$ \textsc{Celerite} terms. For $1 \leq n \leq N$ we have:
\begin{equation}
    a_n = p\theta_n,\,\,
    b_n = q\theta_n,\,\,
    c_n = r, \textrm{ and }
    d_n = s - n\omega_0,
\end{equation}
representing the positive frequencies in the Fourier series, and for $N < n \leq 2N$ we have:
\begin{equation}
    a_n = p\theta_{n-N},\,\,
    b_n = q\theta_{n-N},\,\,
    c_n = r,\textrm{ and }
    d_n = s + (n-N)\omega_0,
\end{equation}
representing the negative frequencies in the Fourier series, and:
\begin{equation}
    a_{2N+1} = p\theta_0,\,\,
    b_{2N+1} = q\theta_0,\,\,
    c_{2N+1} = r\textrm{ and }
    d_{2N+1} = s,
\end{equation}
the central zero value of the Fourier series.
For a $\tau(\delta x)$ with $M$ terms this becomes $M(2N+1)$ terms.

In the simplest case where $\tau(\delta x)$ only has real coefficients, i.e. $s=0$, we can use symmetry to reduce the number of \textsc{Celerite} terms to $N+1$, which may have a notable computational benefit and worth implementing if possible. In this case the terms are
\begin{equation}
    a_n = 2p\theta_n,\,\,
    b_n = 0,\,\,
    c_n = r\textrm{ and }
    d_n = n\omega_0,
    \end{equation}
for $1 \leq n \leq N$, and
\begin{equation}
    a_{N+1} = p\theta_0,\,\,
    b_{N+1} = 0,\,\,
    c_{N+1} = r\textrm{ and }
    d_{N+1} = 0.
\end{equation}
See Section \ref{example_kernel} for a specific example of a kernel that makes use of this optimisation.

In addition to $k(\delta x)$ we also add two white noise terms to the diagonal of the covariance matrix.
The first term represents the measurement noise, as estimated from the variance of the off-pulse noise and is different for data associated with each observation.
For bright pulsars, the `noise' in the profile is usually dominated by random profile variations due to averaging a finite number of pulses, so we also include a single parameter additive white noise term in the GP kernel.

\subsection{A simple example kernel}
\label{example_kernel}
\begin{figure}
    \centering
    \includegraphics[width=8.5cm]{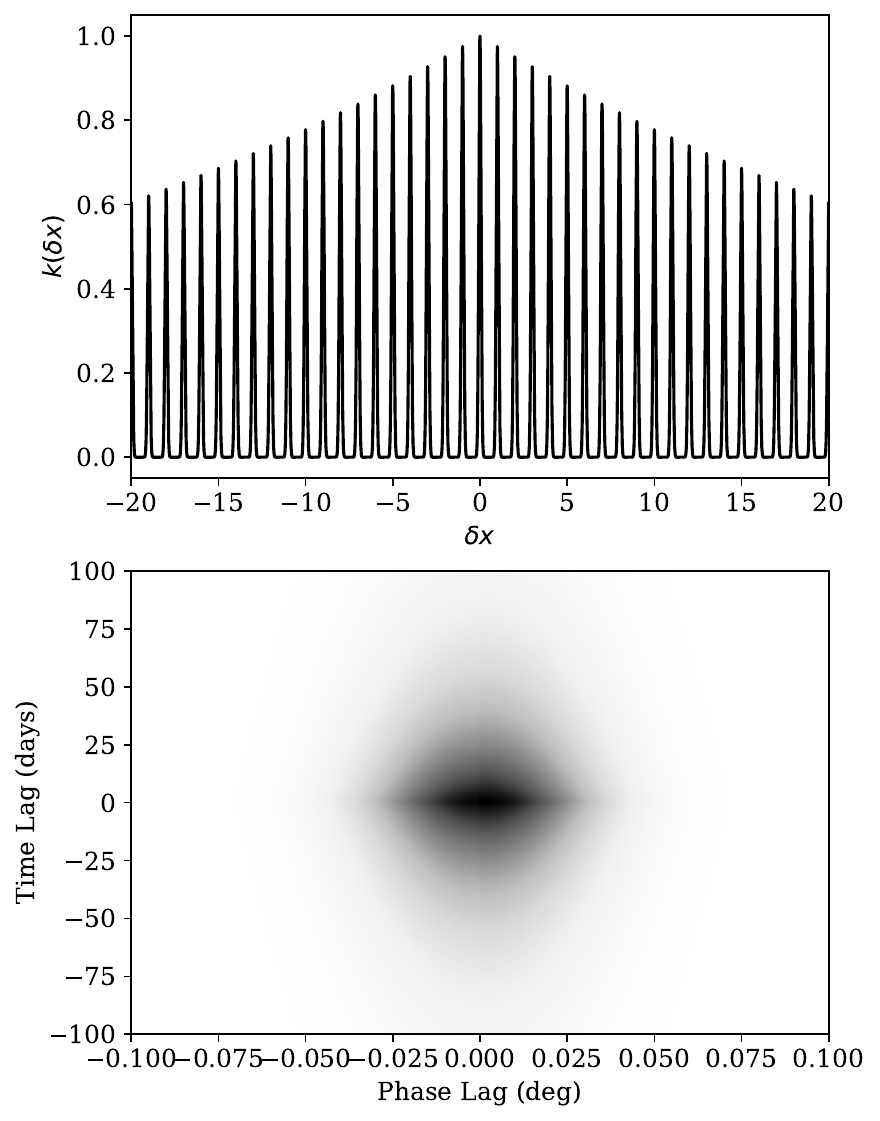}
    \caption{\label{KernelExample}An example of the simple kernel. Left, the 1-D kernel as a function of $\delta x$. Right: The kernel `folded' into 2-D.}
\end{figure}

As there is no a-priori reason to favour one particular kernel for either of our dimensions, we choose as an example to implement a `simple as possible' choice where $\tau(\delta t)$ is an exponential with length-scale $\lambda$ and amplitude $A$, and $\Theta(\delta t)$ is a normalised Gaussian of width $\sigma$. In this case we have $p=A$, $q=0$, $s=0$, and $r=-1/\lambda$.
The Fourier coeficients are 
\[
\theta_n = \sqrt{2\pi\sigma^2}e^{-2\pi \sigma^2 n^2},
\]
i.e. the Fourier transform of a Gaussian.
Since we have defined $x(\phi,T)$ to be periodic on unit length, we choose $\omega_0 = 2\pi$.
Hence, the \textsc{Celerite} coefficients are:
\begin{align}
\begin{split}
    a_n &= 2A\sqrt{2\pi\sigma^2}e^{-2\pi \sigma^2 n^2}, \\
    b_n &= 0, \\
    c_n &= -1/\lambda,\\
    d_n &= 2\pi n,
    \end{split}
\end{align}
for $1 \leq n \leq N$, and a purely real-valued term,
\begin{align}
\begin{split}
    a_{N+1} &= A\sqrt{2\pi\sigma^2}e^{-2\pi \sigma^2 n^2}, \\
    c_{N+1} &= -1/\lambda.\\
    \end{split}
\end{align}
An example of this kernel is shown in Figure~\ref{KernelExample}. The upper panel shows how the kernel is composed of a series of narrow spikes (the $\Theta(\delta t)$ term) which decay exponentially (the $\tau(\delta t)$ term). In the lower panel we `fold' this kernel and convert the axes back to $\phi$ and $T$ to show what the kernel looks in the original dimensions of the data.

The computational efficiency of this kernel trades directly with the quality of the approximation to a Gausisan by adjusting $N$, the number of Fourier terms used.
The required value of $N$ for a good approximation depends strongly on the relative width of the Gaussian and the periodicity, i.e. on $\sigma\omega_0$.
This motivates our choice to define the periodicity to be over the on-pulse window rather than over the entire pulse period, as this can have a very large impact on computational efficiency for pulsars with narrow pulse widths.
The main constraint is that $\sigma\omega_0 \ll 2\pi$ to avoid unintended covariance between the end of one pulse profile and the start of the adjacent one, but it is relatively trivial to add padding should that become an issue.

\subsection{Principle Component Analysis and correlation with {\nudot}}
The GP modelling enables reconstruction of underlying pulse shape changes, even when the individual observations are noise-dominated.
 As in L10, it is also useful to define a `pulse shape parameter' that reduces the reconstructed pulse stack to a 1-D quantity that can be used to measure, for example, the correlation of the pulsar shape with \nudot.
In this paper, we use principle component analysis (PCA) to reduce the dimensionality of the GP output and search for such correlations.
The PCA decomposes the data into a linear combination of eigenvectors, which we term `eigenprofiles' since they are in the form of the pulse profile, and the principal component time series (PCTS) that together reconstruct the original dataset.
Since the PCA is a linear transform of the GP output, they capture the smooth GP reconstruction of the pulse shape variations.
It is important to note that, as with any PCA analysis, the eigenprofiles only capture the variance in the underlying dataset, and emerging structures do not necessarily have a direct physical interpretation.
Whilst care is needed not to overinterpret their meaning, they provide a purely statistical impression of the variations, free from biases that may be present in other modelling (e.g. by hand selecting particular shape parameters).

Correlations between the PCTS and \nudot\ are computed by evaluating the GP reconstructed profile at each observation, applying the eigenprofile decomposition, and computing the Pearson $r$ correlation coefficient with \nudot\ at the same epochs.
The calculation of the error on $r$ is complicated by the intrinsically correlated nature of the reconstructed samples, effectively reducing the number of measurements that contribute meaningfully to the estimate of $r$.
To account for this, we estimate the error on $r$ by drawing 1000 realisations of the maximum likelihood GP model and computing the variance of the resulting correlation coefficients of these with {\nudot}.

\subsection{Psrcelery}
We have implemented this algorithm in a package named \textsc{psrcelery}, which is openly available for use\footnote{\url{https://github.com/SixByNine/psrcelery}}.
As of publication, \textsc{psrcelery} includes the `simple as possible' model of an exponential for $\tau(\delta x)$, and a Gaussian kernel for $\Theta(\delta x)$, as well as the ability to use arbitrary single-term \textsc{Celerite} kernels for $\tau(\delta x)$.
In the results of this paper we make use of the widely used ``Mat\'ern-3/2" kernel for $\tau(\delta x)$, with length scale $\lambda$ and amplitude $A$, and defined by
\begin{equation}
    k({\tau}) = A \left(1+
        \frac{\sqrt{3}\,\tau}{\lambda}\right)\,
        \exp\left(-\frac{\sqrt{3}\,\tau}{\lambda}\right),
\end{equation}
which can be arbitrarily well approximated using a single-term \textsc{Celerite} kernel~\citep{celerite}.

\section{Data}
In this paper we apply \textsc{psrcelery} to data from the 76-m Lovell telescope at Jodrell Bank Observatory.
We use extended versions of the datasets presented in S22, focusing on the $\sim 14$ years of data from the DFB backend recorded since 2009.
These data are primarily recorded at a centre frequency of 1520 MHz, using a 384~MHz bandwidth, with a small number of early observations at a centre frequency of 1400 MHz and a 128 MHz bandwidth.
The data are all folded in 10s sub-integrations with 768 frequency channels and 1024 phase bins per pulse period.
Radio frequency interference is mitigated by median filtering, followed by inspection and deletion of affected frequency channels and sub-integrations.
The resulting observations are then averaged in time and frequency (after correction for interstellar dispersion) to produce a single pulse profile per observation.
The profiles are aligned and scaled by cross-matching with a noise-free template (see \citealp{taylor92}) and rotated to alignment by applying phase shifts to the Fourier transform.
The on and off pulse regions are selected manually, and the off-pulse rms is computed to identify and remove any observations with spurious data from e.g. unmitigated radio frequency interference.

In some cases, we also extend the dataset to earlier times by appending data from the earlier analogue filterbanks (AFB), which used a 32 MHz band centred at 1400 MHz and 400 phase bins per pulse period.
For the combined data, the DFB data are resampled by means of Fourier transform to give the same phase resolution as the AFB data, and are then processed as a single consistent dataset.
We note that we did not apply this broadly to all pulsars because in many cases the change in overall quality of the profiles between AFB and DFB dominates the subsequent analysis, and because the AFB dataset is identical to that already studied in depth in S22 and L10.

\subsection{Timing}
In order to compare the observed pulse shape variations with \nudot, we form arrival times for each observation and derive the {\nudot} from the pulsar timing model using the method described in \citet{keith23}.
Here, a Fourier-domain GP model for the excess spin noise is fit to the dataset simultaneously with the pulsar timing model.
A time-series of \nudot\ is then computed analytically from the maximum-likelihood solution of the spin noise.
This differs slightly from the approach of S22, where the timing model was fit first, and a time-domain GP was fit to the residuals.
Here, as demonstrated in \citet{keith23}, we avoid the risk of errors in the timing model affecting the GP model by fitting for the pulsar timing parameters and GP model simultaneously.

\section{Results}
\label{results}
The input to the model are all pulse profiles, aligned and normalised using a standard template, and with the median of each phase bin subtracted to give a `difference map' of the deviations of the pulsar signal from the typical pulse profile.
The results from \textsc{psrcelery} are presented in the form of a multi-panel summary figure for each pulsar, see e.g. Figure \ref{rainbow0740}.
These figures show the difference map as reconstructed from the GP analysis, as well as the results of the PCA analysis.
The formal error in the GP reconstruction is used as a transparency mask on the figures such that colours reach full intensity where the GP reconstruction is more than 3-$\sigma$ from zero.
However, we caution that this formal error is likely underestimated, and excess noise can be seen in the leading and trailing edges of the profile.
As these figures are primarily for visualisation, we leave this excess noise in order to guide the eye as to the inherent noise levels.

\begin{table}
\caption{\label{results_table} Observed correlation between pulse shape and \nudot\ for the sample of 21 pulsars with profile shape variations. Shape classifications `M' and `W' refer to relative enhancement of the wings or peak respectively (see Section \ref{results}), and `C' indicates a more complex shape change. A `+' symbol indicates shape variations in addition to that in the classification. A `?' symbol indicates that the classification is doubtful. The Pearson correlation coefficient, $r$, is given as an indication of the correlation between the PCTS and \nudot. The first eigenprofile is used for the correlation, except for those marked with~*, specifically PSR B1826$-$17 uses the fourth eigenprofile and B0611+22, B0919+06 and J2043+2740 use the sum of the first two eigenprofiles. The fourth column is a comment on the apparent correlation of {\nudot} and the profile shape, see the relevant section for a full discussion. We also indicate if correlations were identified in previous works, L10, S22, N22 \citep{nitu22} and B24 \citep{basu24}. Footnotes after the pulsar name indicate references to studies of each pulsar at a range of observing frequencies.}
\begin{tabular}{llllll}
PSR& Shape &  $r$ & Corr. & Prev. Work & Fig.\\
\hline
B0105+65        \textsuperscript{a\,c}\kern-1em & M   & $0.47 \pm 0.12$ & Yes                  & --                   & \ref{rainbow0105}   \\
B0144+59        \textsuperscript{}\kern-1em & M   & $0.53 \pm 0.11$ & Yes                  & Yes (N22)            & \ref{rainbow0144}   \\
B0611+22        \textsuperscript{a\,b\,d}\kern-1em & M+   & $0.57 \pm 0.11$ & Yes*                 & --                   & \ref{rainbow0611}   \\
B0626+24        \textsuperscript{a\,c\,d}\kern-1em & W   & $0.13 \pm 0.14$ & Partial              & --                   & \ref{rainbow0626}   \\
B0740$-$28      \textsuperscript{b\,e\,h}\kern-1em & M/C   & $0.40 \pm 0.11$ & Yes                  & Yes (L10)            & \ref{rainbow0740}   \\
B0919+06        \textsuperscript{a\,d}\kern-1em & W   & $0.55 \pm 0.14$ & Yes*                 & Yes (L10)            & \ref{rainbow0919}   \\
B0950+08        \textsuperscript{b\,d}\kern-1em & M?   & $0.09 \pm 0.10$ & No                   & No (S22)             & \ref{rainbow0950}   \\
B1540$-$06      \textsuperscript{a\,b}\kern-1em & M   & $0.73 \pm 0.22$ & Yes                  & Yes (L10)            & \ref{rainbow1540}   \\
B1642$-$03      \textsuperscript{a\,b}\kern-1em & M+   & $0.77 \pm 0.20$ & Yes                  & Yes (S22)            & \ref{rainbow1642}   \\
J1740+1000      \textsuperscript{b\,c\,d}\kern-1em & ?   & $0.08 \pm 0.13$ & Cmplx                & --                   & \ref{rainbowJ1740}  \\
B1740$-$03      \textsuperscript{}\kern-1em & W   & $0.14 \pm 0.10$ & Partial              & --                   & \ref{rainbow1740}   \\
B1818$-$04      \textsuperscript{a\,b}\kern-1em & M/W   & $0.22 \pm 0.17$ & No                   & No (S22)             & \ref{rainbow1818}   \\
B1822$-$09      \textsuperscript{a\,b\,e}\kern-1em & M+   & $0.19 \pm 0.10$ & Cmplx                & Yes (L10)            & \ref{rainbow1822}   \\
B1826$-$17      \textsuperscript{b}\kern-1em & C   & $0.71 \pm 0.19$ & Yes*                 & No (S22)             & \ref{rainbow1826}   \\
B1828$-$11      \textsuperscript{b}\kern-1em & M   & $0.81 \pm 0.12$ & Yes                  & Yes (L10)            & \ref{rainbow1828}   \\
B1842+14        \textsuperscript{a\,b\,c\,d}\kern-1em & M?   & $0.23 \pm 0.22$ & Partial              & Yes (B24)            & \ref{rainbow1842}   \\
B1914+09        \textsuperscript{a\,b}\kern-1em & M   & $0.44 \pm 0.19$ & Yes                  & Yes (B24)            & \ref{rainbow1914}   \\
B1917+00        \textsuperscript{a\,f}\kern-1em & W   & $0.73 \pm 0.19$ & Unclear              & No (B24)             & \ref{rainbow1917}   \\
B2035+36        \textsuperscript{a\,c\,f}\kern-1em & M+   & $0.93 \pm 0.24$ & Yes                  & Yes (L10)            & \ref{rainbow2035}   \\
J2043+2740      \textsuperscript{c\,g}\kern-1em & W+   & $0.59 \pm 0.16$ & Partial*             & Yes (L10)            & \ref{rainbow2043}   \\
B2148+63        \textsuperscript{a\,c}\kern-1em & C   & $0.37 \pm 0.18$ & Partial              & No (S22)             & \ref{rainbow2148}   \\

\hline
\end{tabular}
\textsuperscript{a}\,\citet{PHS15}, 
\textsuperscript{b}\,\citet{jk18}, 
\textsuperscript{c}\,\citet{BKK}, 
\textsuperscript{d}\,\citet{Olszanski19}, 
\textsuperscript{e}\,\citet{KJ06}, 
\textsuperscript{f}\,\citet{RWVO22}, 
\textsuperscript{g}\,\citet{WRVO}, 
\textsuperscript{h}\,\citet{JK06}, 
\end{table}

We also attempt to determine correlations between the profile shape and \nudot, by looking at the correlation between the PCTS from the first eigenprofile and the measured \nudot\ timeseries.
We compute the Pearson correlation coefficient between the PCTS and {\nudot} as estimated at each observation, in order to avoid biasing the correlation computation during periods where the pulsar is not regularly observed.
The eigenprofiles (and PCTS) from the PCA have an arbitrary sign, hence we invert the sign of the eigenprofile and PCTS if the measured correlation is negative.
This ensures that the eigenprofiles as presented represent the shape change most associated with an increase in {\nudot} (or a decrease in $|\dot{\nu}|$), and the PCTS always have a positive correlation, which is easier to assess in the plots.

Table \ref{results_table} shows a summary of these results, as well as qualitative assessments of the correlation and shape of the profile changes.
For each pulsar, we also attempt to classify the shape of the eigenprofile that captures the variation most correlated with \nudot.
In the clearest examples, this is in the form of a relative change between the intensity of the centre and the edges of the profile.
We class those with an increase in the edges and a decrease in the centre as `M' shape, and in the converse case (increased centre, decreased edges) we classify it as `W' shape.
Where the eigenprofile is more complex with multiple inversions such that it cannot classified into either `M' or `W' shape we label it `C'.
Table \ref{results_table} also includes a qualitative assessment of the correlation of the PCTS with \nudot.
These pulsars all have high-quality observations across observing frequency and in full polarisation available from the EPN profile database (\url{https://psrweb.jb.man.ac.uk/epndb}), and we refer readers to the papers indicated by the footnotes in Table \ref{results_table} for additional context. A broader discussion of the geometry and emission of these pulsars can be found in \citet{Olszanski19} and \citet{2022MNRAS.514.3202R}.

\subsection{The Lyne et al. (2010) sample}
Here we look at the 17 pulsars that appear in L10 and revisited in S22 and the \nudot\ timeseries studied in \citet{keith23}.
\subsubsection{PSR B0740$-$28}
\begin{figure}
    \centering
    \includegraphics[width=8.5cm]{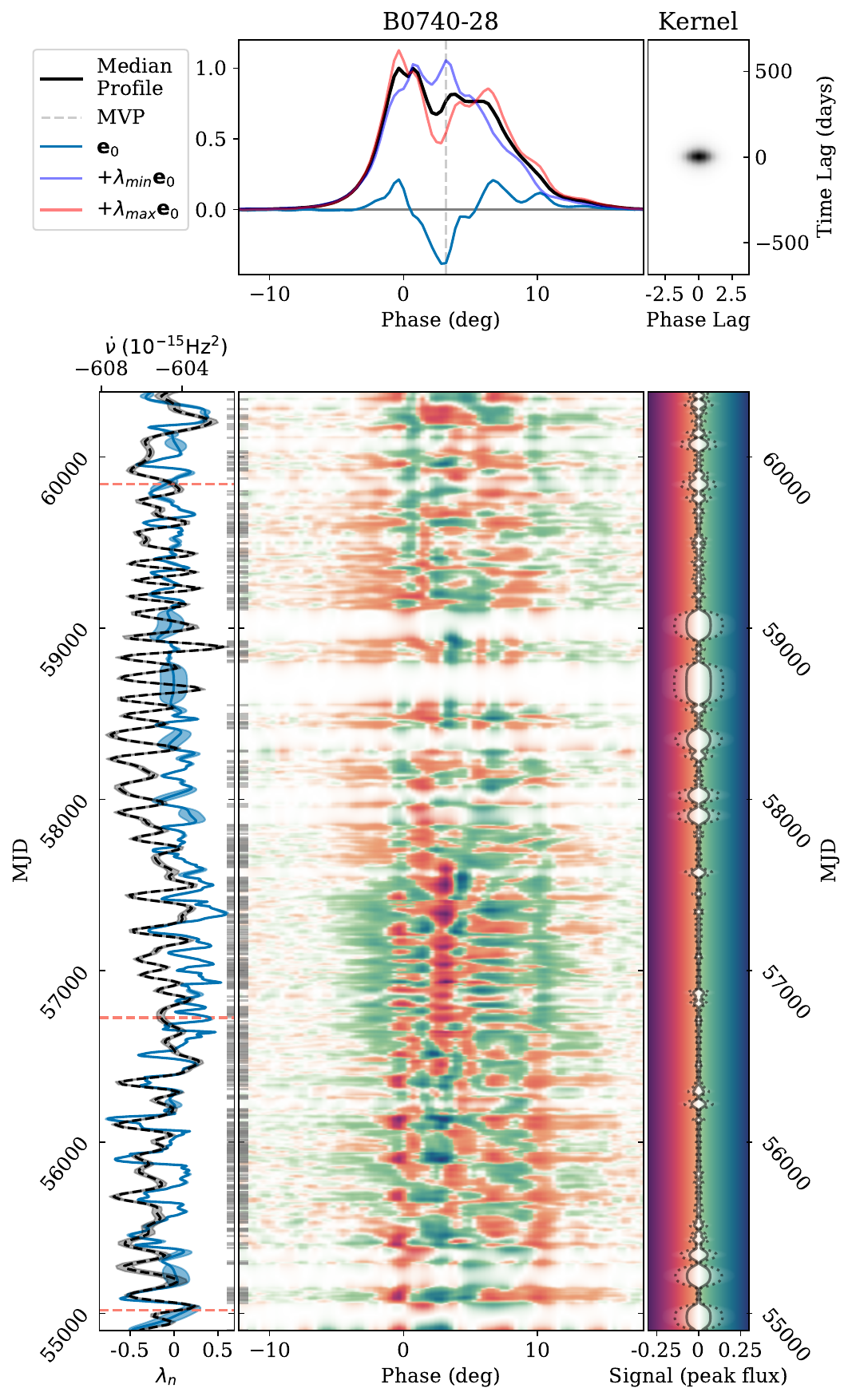}
    \caption{\label{rainbow0740}The output of \textsc{psrcelery} for PSR B0740$-$28.
    The central panel shows the GP reconstructed deviation from the median profile as a function of MJD and phase. Short ticks mark the epoch of each observation.
 The colour map is adjusted in each row to fade to white where the uncertainty in the model prediction drops below 3-$\sigma$.
 The right-hand panel shows the colour-scale of each row, and the solid and dotted lines show the 1 and 2-$\sigma$ formal uncertainty on the GP reconstruction.
The left panel shows the PCTS associated with the first eigenprofile of the PCA analysis (solid coloured line), overlaid with the GP-derived model of the \nudot\ time-series (black dashed line). The value of \nudot\ increases towards the right, hence a higher spin-down rate is to the left. The epoch of glitches are shown by dashed horizontal lines.
The central upper panel shows the median pulse profile (thick black line), the first eigenprofile and the sum of the median profile and the first eigenprofile multiplied by the minimum (blue) and maximum (red) of the PCTS, in order to give an impression of the scale of the profile changes. The dashed gray line marks the phase bin with the most variability (MVP). The upper-right panel shows the size and shape of the GP kernel, on the same scale as the main panel.}
\end{figure}

PSR B0740$-$28 is known to have complex and time-variable quasi-periodic pulse shape variation that clearly correlates at times with \nudot, but can also appear anti-correlated at other times (\citealp{keith13}, L10, S22, \citealp{0740_IAR}). Figure~\ref{rainbow0740} shows the reconstructed profile evolution of PSR B0740$-$28 from the GP model.
The overall impression from the profile evolution is that the quasi-periodic profile variations change in both shape and periodicity, with the oscillations remaining quasi-stable for hundreds to thousands of days before seemingly transitioning within a few \nudot\ cycles to a new state.
The eignenvalues associated with the first eigenprofile have a correlation with \nudot\ of $r=0.4\pm0.1$ over our observing window, however the correlation is clearly much better at some epochs than others.
For example, between the first two glitches (dashed pink lines in Figure \ref{rainbow0740}) the correlation is very clear, but almost inverts by MJD 58000.
\begin{figure}
    \centering
    \includegraphics[width=8.5cm]{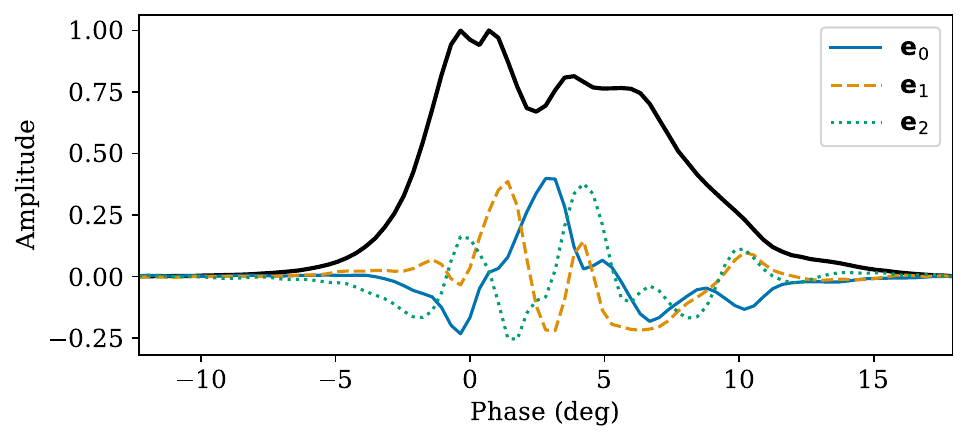}\\
    \includegraphics[width=8.5cm]{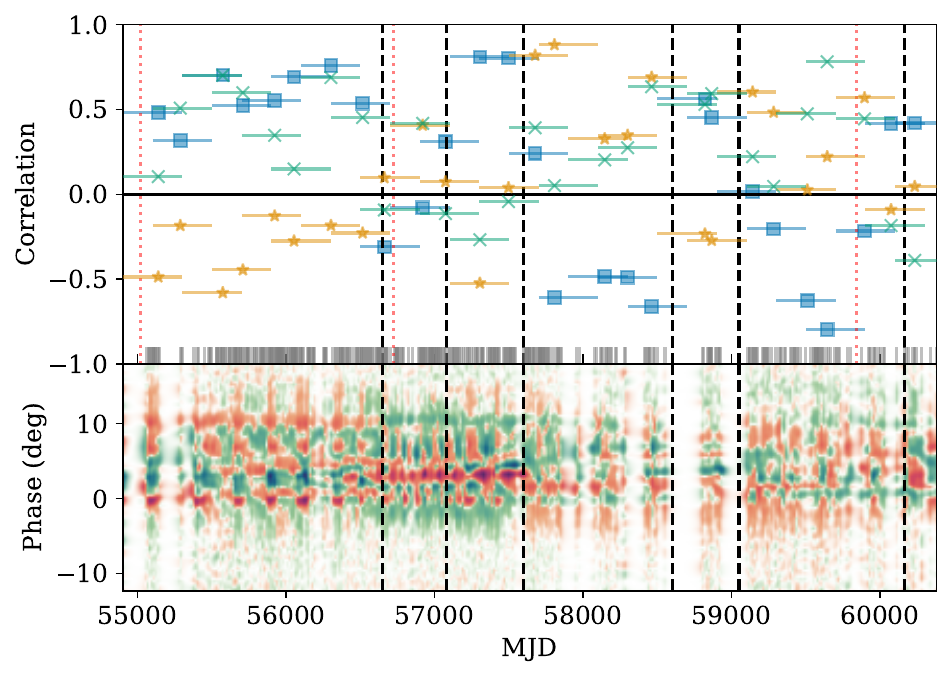}
    \caption{\label{0740_correlation} Upper panel: The first three eigenprofiles for PSR B1740$-$28. Middle panel: Correlation of the first three eigenprofiles (square, star and cross markers) of PSR B1740$-$28 with \nudot. Correlations were computed over sliding 400-day windows, stepping by 200 days each time. Vertical dashed lines indicate the different regions analysed in Figure \ref{0740_evolution}. Lower Panel: The output of the GP reconstruction as in Figure \ref{rainbow0740}.}
\end{figure}
This is seen more clearly in Figure \ref{0740_correlation}, where we have correlated the PCTS from the first three eigenprofiles with \nudot\ over sliding 400-day windows.
Each eigenprofile leads to a correlation with \nudot\ at some epochs, but can also strongly negatively correlate at other times.
The GP-reconstructed pulse stack seems to show some distinct changes of `character' of the shape changes, and these also loosely agree with the correlation between \nudot\ and the various PCTS.
\begin{figure*}
    \centering

        \includegraphics[width=17cm]{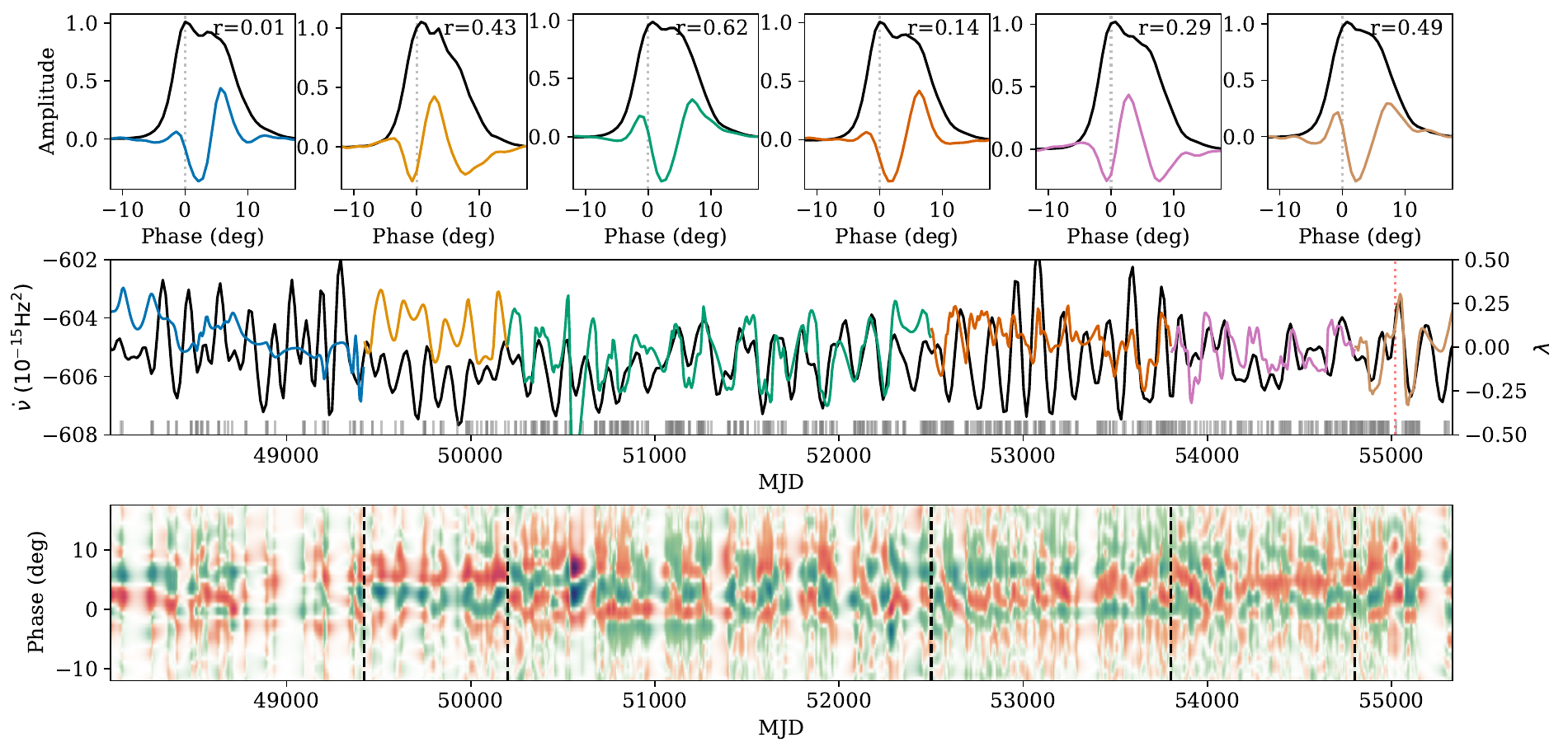}   \\
        \includegraphics[width=17cm]{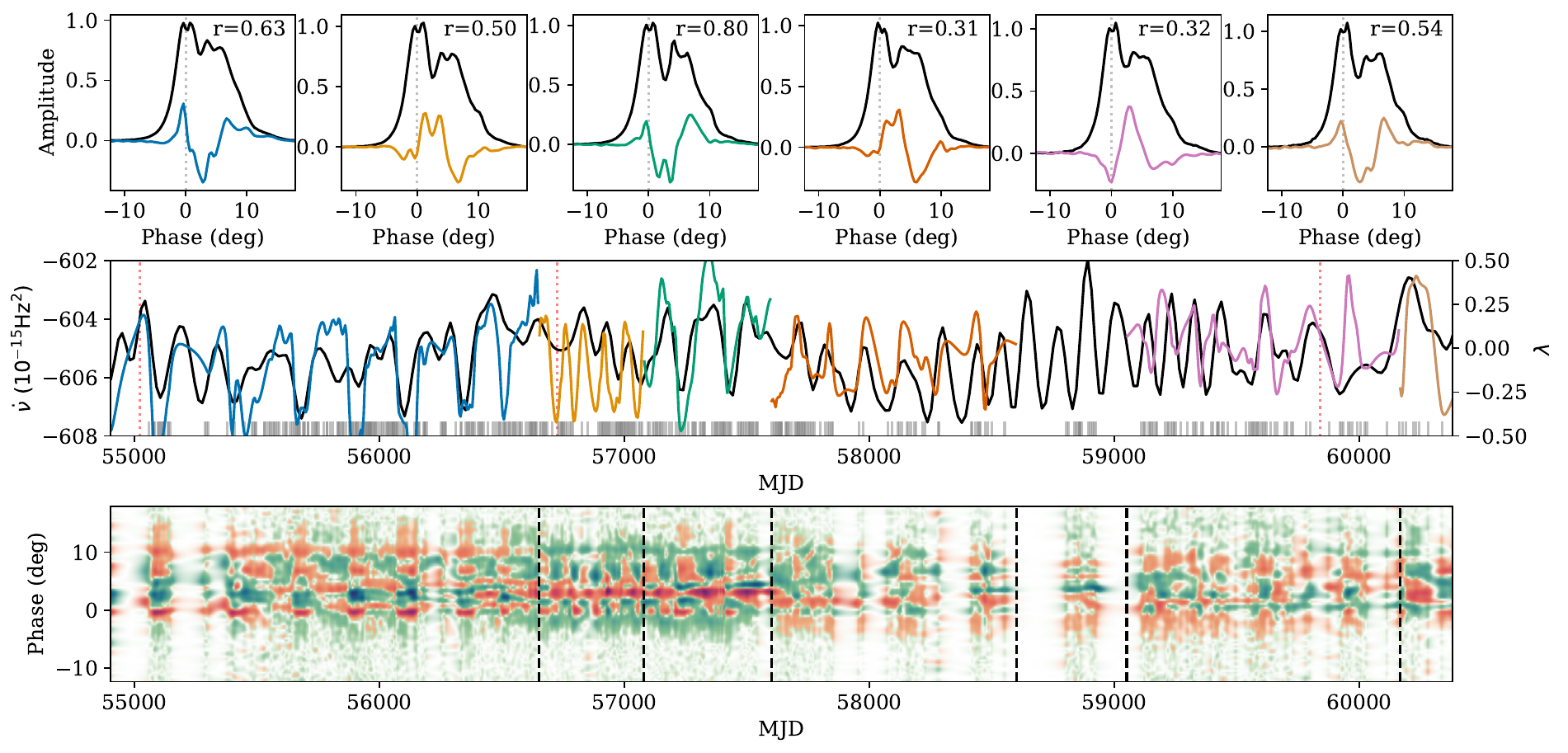}  
    
    \caption{\label{0740_evolution}PCA analysis performed on short segments of the Gaussian process model for PSR B0740$-$28. Upper three rows show AFB data and lower three rows show DFB data. The upper row shows the median profile within each segment, and the first eigenprofile from the PCA analysis. The middle row shows the \nudot\ timeseries overlaid with the PCTS from each segment. The lower panel shows the Gaussian process output. Vertical dashed lines show the boundaries of the segments. Vertical dotted lines indicate the epoch of glitches.}

    \end{figure*}
To investigate this further, we broke the pulse stack into short segments as indicated by the black dashed lines on Figure \ref{0740_evolution}, and performed the PCA analysis independently on each segment.
In the lower panel of Figure \ref{0740_evolution} (covering the DFB observations) the correlation between \nudot\ and the first PCTS computed from each individual segment is good, but the shape of the corresponding eignprofile changes significantly between segments.
The eigenprofiles have each been adjusted in sign so that they correspond to a positive correlation with \nudot.
Note that profiles are only constructed from observations in the 21-cm band in order to avoid complications of shape variation with observing frequency, but the the \nudot\ time series are derived from the full timing model that includes other observations at lower frequencies and including daily observations with the 42ft telescope since MJD $\sim52200$.
This is apparent, for example, in the fifth segment of the lower panel where we have insufficient profile data for the PCA analysis to be useful, but are still able to measure \nudot\ from the daily observations with the 42ft telescope.
During the first, third and final segments of the DFB data, where the \nudot\ variation appears slowest, the profile shape variation has a positive dependence on the height of the peak close to phase zero, a negative dependence in the centre and a generally positive dependence on the trailing shoulder.
The three other regions (second, fourth and sixth), where the \nudot\ variations are oscillating more quickly, seem to show an almost inverted dependence, and the overall correlation seems typically weaker.
We repeat the analysis on the AFB dataset, shown in the upper half of Figure \ref{0740_evolution}.
Here, broadly similar features are observed, however the correlation between the PCTS and \nudot\ is not always clear, especially in the first chunk where the profile observations seem insufficiently dense to properly capture the rapid profile variations.

\citet{keith13} suggested that the correlation coefficient changed around the time of the moderate sized glitch ($\Delta\nu/\nu \sim 10^{-7}$) at MJD 55022. The pulsar has seen two subsequent glitches, a relatively small ($\Delta\nu/\nu \sim 10^{-9}$) glitch around MJD 56726 and a large glitch ($\Delta\nu/\nu \sim 4\times10^{-6}$) at MJD 59839, which has also been suggested to be synchronous with a change in the profile shape and spin-down variations \citep{0740_IAR}.
The profile shape variations also appear to change close to the glitch at MJD 56726, though it should be noted, as with the MJD 55022 glitch, that the change in the profile seems to predate the glitch by at least 100 days.
The MJD 59839 glitch occurs during a time where the observing coverage was fairly sparse, but we find that there does not seem to be any obvious change in the profile variations until about a year after the glitch.
Given the frequent changes in the expression of the profile variations, it seems plausible that the apparent relation between glitch activity and emission may be coincidental, especially as there are several transitions that are not associated with any measured glitch activity.

\subsubsection{PSR B0919+06}

\begin{figure}
    \centering
    \includegraphics[width=8.5cm]{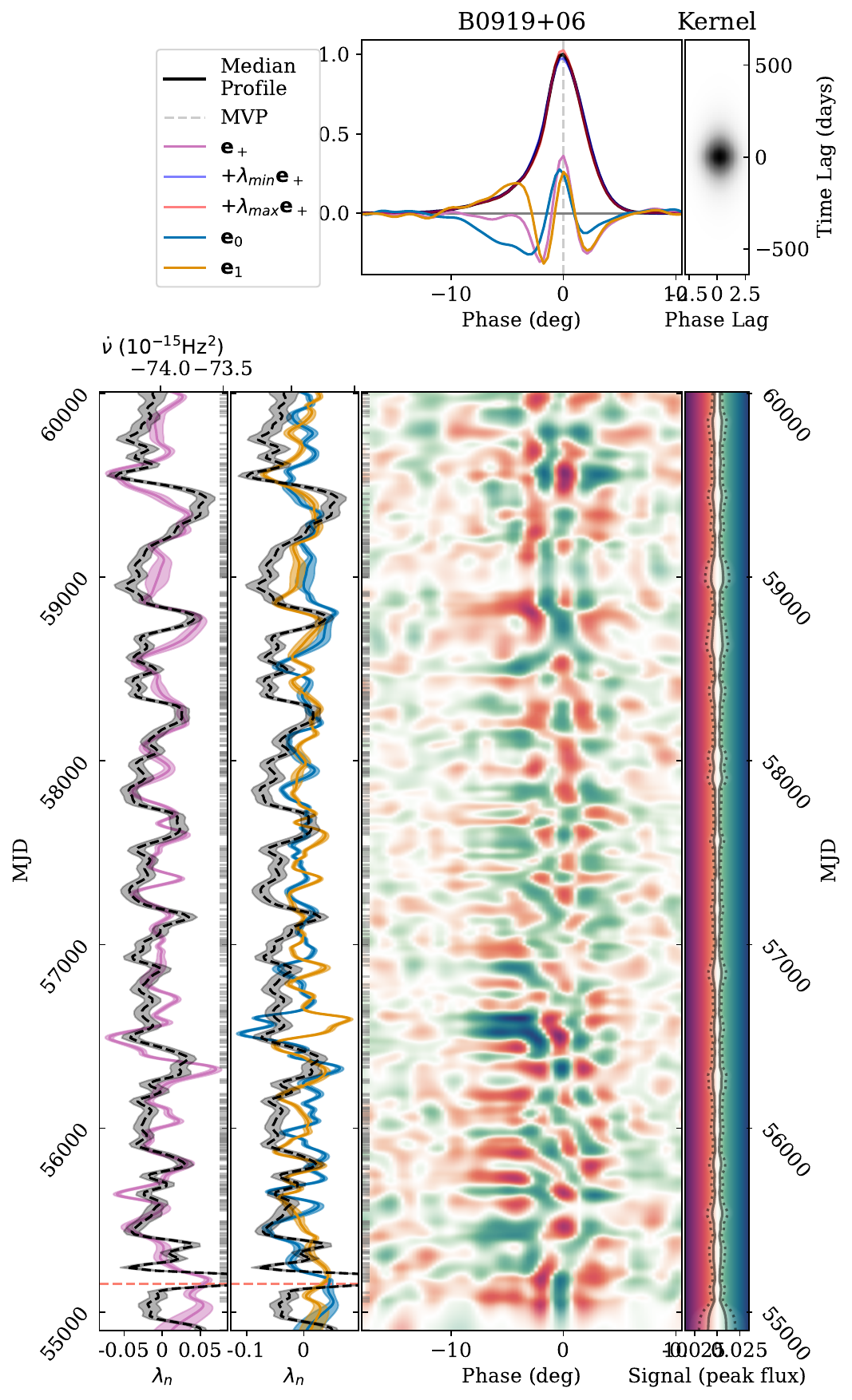}
    \caption{\label{rainbow0919}The output of \textsc{psrcelery} for PSR B0919$+$06. As Figure~\ref{rainbow0740}, except an additional leftmost panel shows the correlation of the PCTS computed from the sum of the first two eigenprofiles.}
\end{figure}

PSR B0919+06 is known to exhibit `flares' where the emission appears shifted earlier in phase for a few tens of pulses every few thousand pulses.
This has been shown to be unrelated to the \nudot\ evolution, which is better correlated with longer-term subtle variations in the pulse shape \citep{psw+14}.
Figure \ref{rainbow0919} shows the \textsc{psrcelery} output for PSR B0919+06.
We find a weak correlation, 0.34, of the first eigenprofile with the \nudot, and a slightly stronger correlation, 0.46, with the second eigenprofile.
However, inspection of the time-series shows that neither set of principle components track the \nudot\ timeseries well in places, likely because of the influence of strong flaring in some epochs.
The first eigenprofile looks 'W' shaped, and the second eigenprofile is similar except for an additional change of sign in the leading edge.
We can improve the correlation with \nudot\ by forming the sum of these two eigenprofiles, $\mathbf{e}_+ = (\mathbf{e}_0+\mathbf{e}_1)/\sqrt{2}$, shown as a pink line on Figure \ref{rainbow0919}. This has the effect of cancelling out the leading shoulder where the `flares' occur.
The PCTS computed from $\mathbf{e}_+$ correlate well with {\nudot}, $r=0.55\pm0.14$, and the associated profile change associated is clearly `W'-shaped.

\begin{figure}
    \centering
    \includegraphics[width=8.5cm]{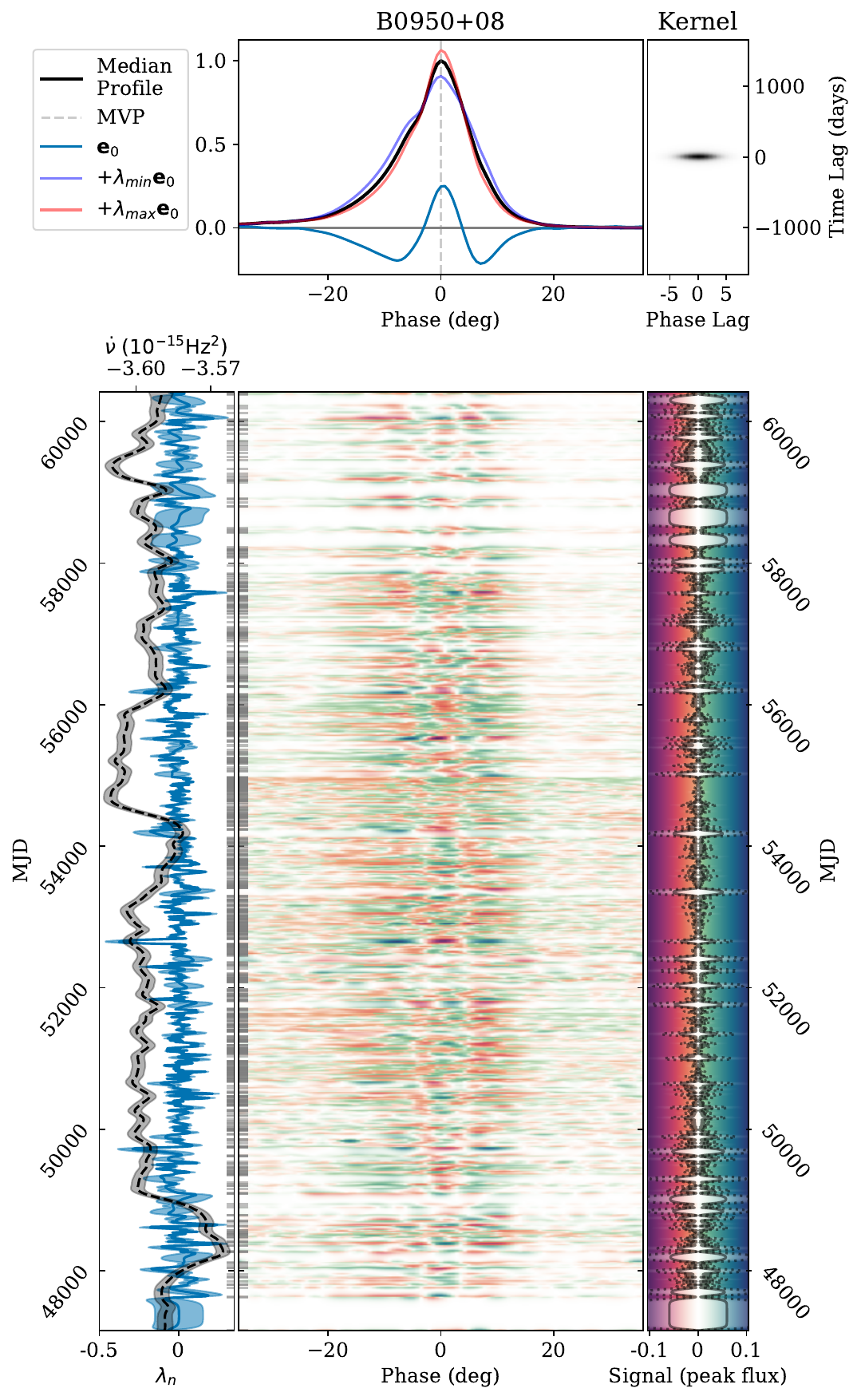}
    \caption{\label{rainbow0950}The output of \textsc{psrcelery} for combined AFB and DFB datasets for the main pulse of PSR B0950+08. Otherwise as Figure~\ref{rainbow0740}.}
\end{figure}

\subsubsection{PSR B0950+08}
S22 did not report any pulse shape variations in PSR B0950+08, and our analysis (Figure \ref{rainbow0950}) suggests that the pulsar exhibits variations on timescale similar to the typical observing cadence, though there do seem to be some occasional instances where the same profile shape persists over multiple subsequent observations.
However, since the \nudot\ timeseries for this pulsar evolves on much longer timescales there is no correlation found between the pulse shape and \nudot.
The first eigenprofile is of the form of an `M' or `W' shape, which are only distinguished by the sign of the correlation with \nudot, and hence equivalent in this case.
We note that if the spin-down rate was varying on the timescale of the pulse shape changes, then this would be faster than we could detect and hence our observed \nudot\ timeseries may reflect a slower time-average of the underlying spin-down state.

\begin{figure}
    \centering
    \includegraphics[width=8.5cm]{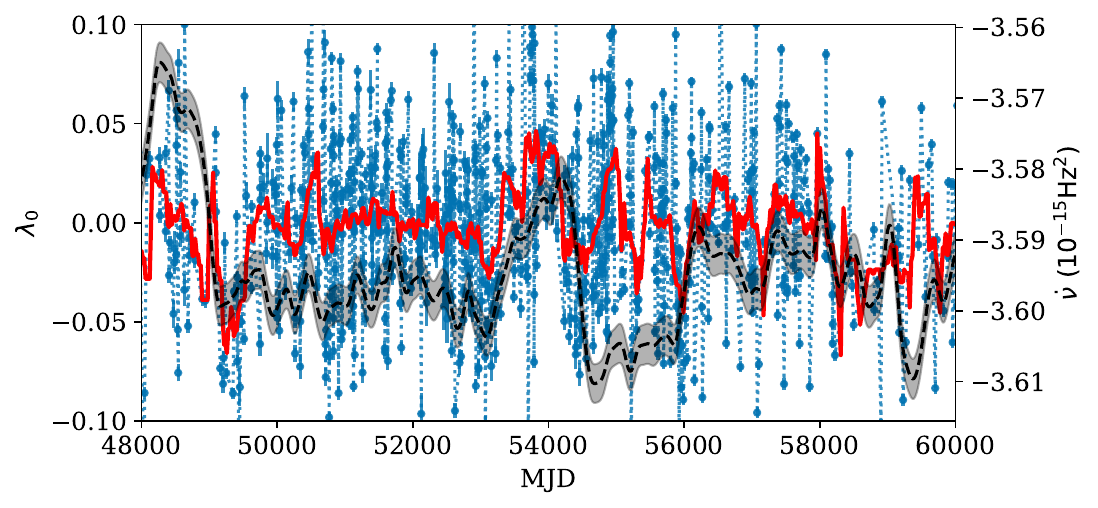}
    \caption{\label{zoom0950} The \nudot\ and PCTS of the first eigenprofile ($\lambda_0$) for PSR B0950+08. \nudot\ is shown as the black dashed line. The PCTS shown as points is computed directly at each observation epoch to more clearly show the rapid variability and the solid line shows the running mean over 200 day windows. \vspace{-1em}}
\end{figure}

In order to see if there is any change in the average profile shape during different \nudot\ states, we compute a running mean of the PCTS over a sliding 200 day window, shown overlaid with the PCTS and the \nudot\ timeseries in Figure \ref{zoom0950}.
Interestingly there do seem to be some structures in the averaged PCTS, and there is some hint of correlation with \nudot\ for some of these.
However the overall correlation is very weak ($r\sim0.1$) and if there is a correlation between profile shape and \nudot\ it is not easily picked up with our analysis.
If we take this correlation for the basis of identifying the sign of the eigenprofile, then it has the same `M' shape as shown in Figure \ref{rainbow0950}.

\begin{figure}
    \centering
    \includegraphics[width=8.5cm]{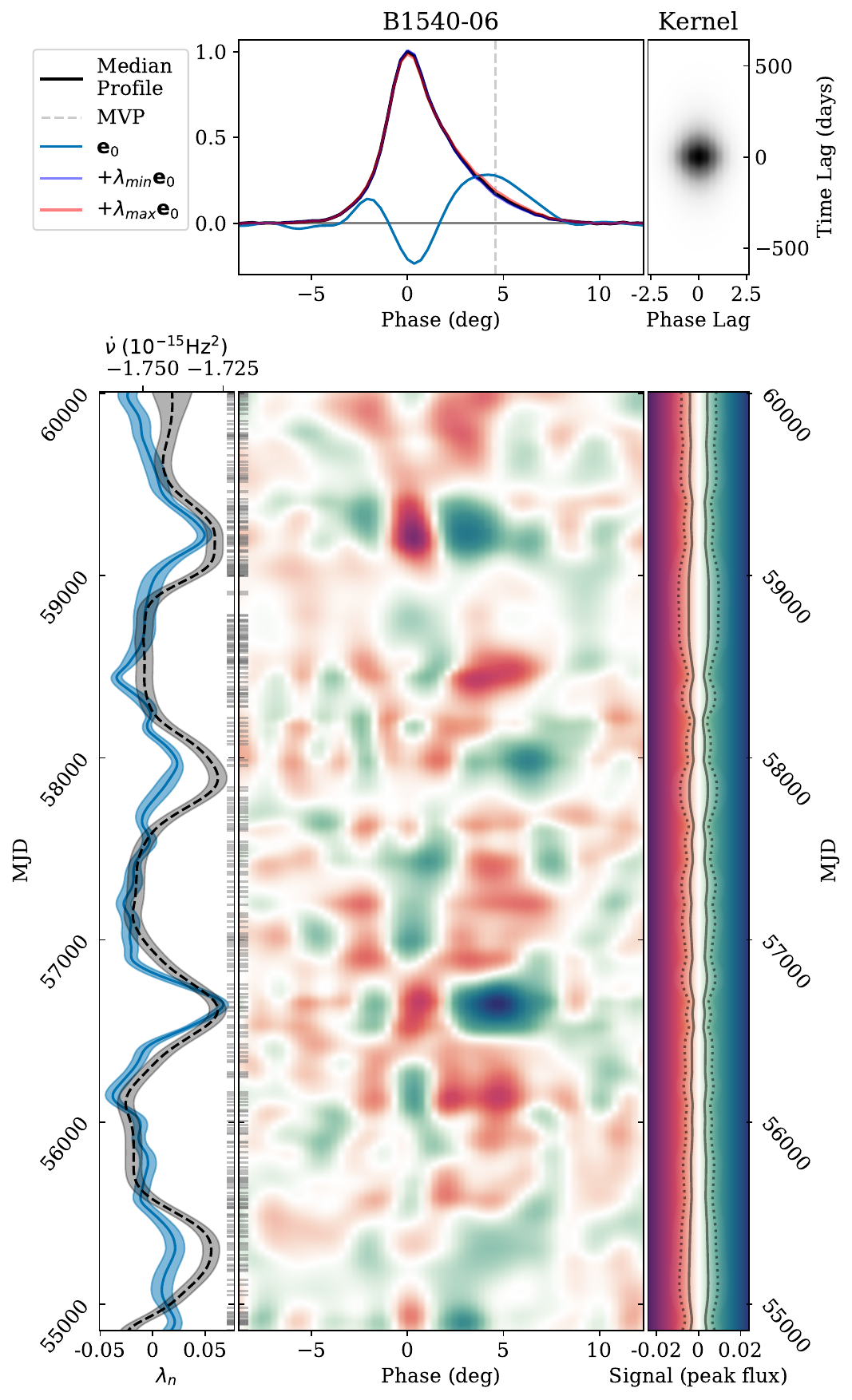}
    \caption{\label{rainbow1540}The output of \textsc{psrcelery} for PSR B1540$-$06. Otherwise as Figure~\ref{rainbow0740}.}
\end{figure}

\subsubsection{PSR B1540$-$06}
Figure \ref{rainbow1540} shows the \textsc{psrcelery} output for PSR B1540$-$06.
This pulsar exhibits a highly periodic \nudot\ timeseries, which correlates very well with profile shape (L10,S22).
Here we also find a strong correlation ($r=0.7\pm0.2$) between \nudot\ and the first eigenprofile.
The profile variations are a very subtle change in the relative amplitude of the peak and shoulders of the profile, highlighting that the GP analysis is effective at identifying such time and phase correlated changes even when the magnitude is very small.
The eigenprofile is generally `M' shaped, though not symmetric.

\subsubsection{PSR B1642$-$03}
\begin{figure}
    \centering
    \includegraphics[width=8.5cm]{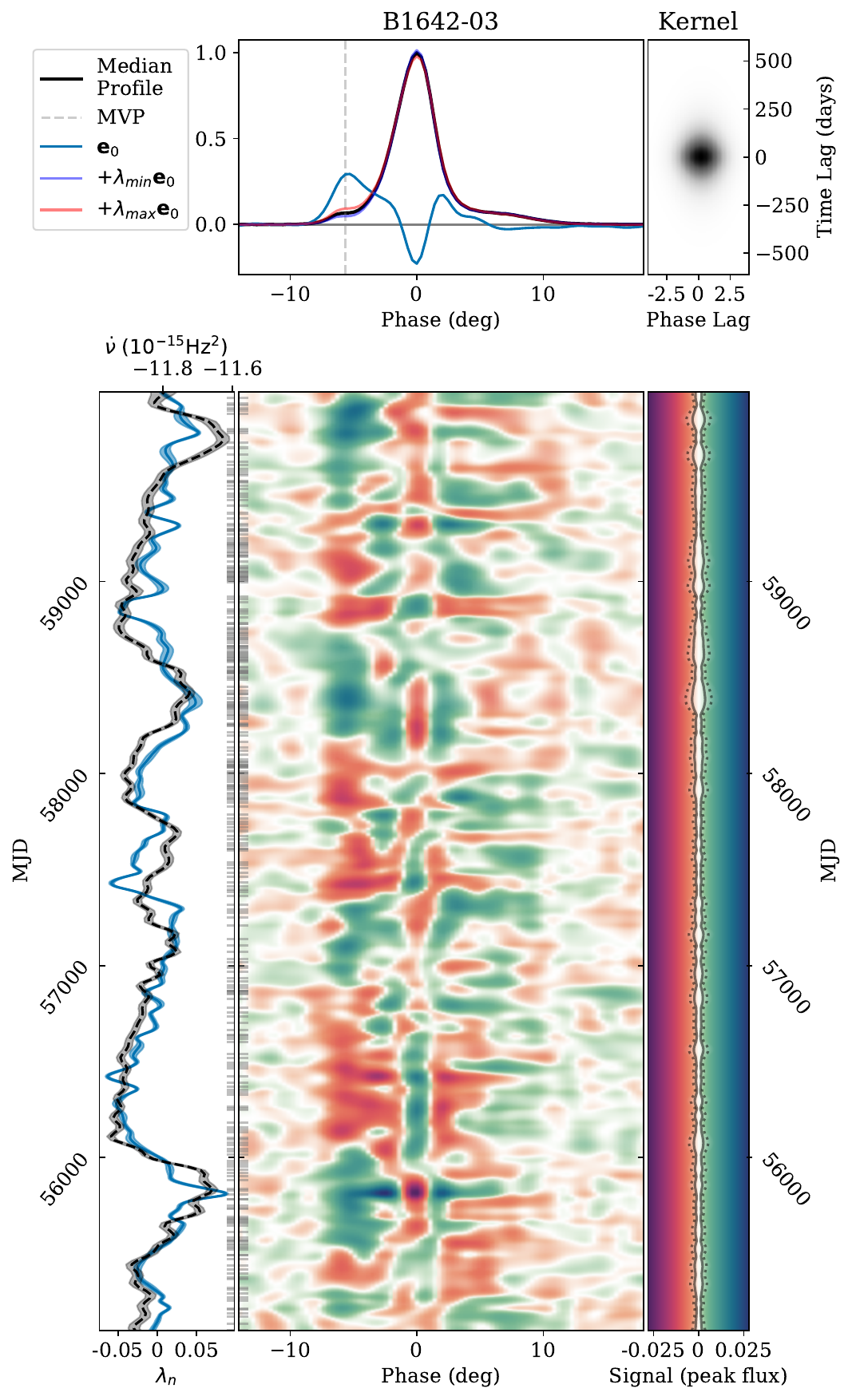}
    \caption{\label{rainbow1642}The output of \textsc{psrcelery} for PSR B1642$-$03. Otherwise as Figure~\ref{rainbow0740}.}
\end{figure}

PSR B1642$-$03 was first identified to have correlated profile shape and \nudot\ in S22, and the \textsc{psrcelery} output is shown in Figure \ref{rainbow1642}.
We confirm that there is a clear correlation between profile shape and \nudot\ from our analysis.
The correlation between the \nudot\ timeseries and the PCTS is $r=0.8\pm0.2$ and the eigenprofile has an `M' shape with an emphasis on the growth of the small leading edge component.

\subsubsection{PSR B1818$-$04}
\begin{figure}
    \centering
    \includegraphics[width=8.5cm]{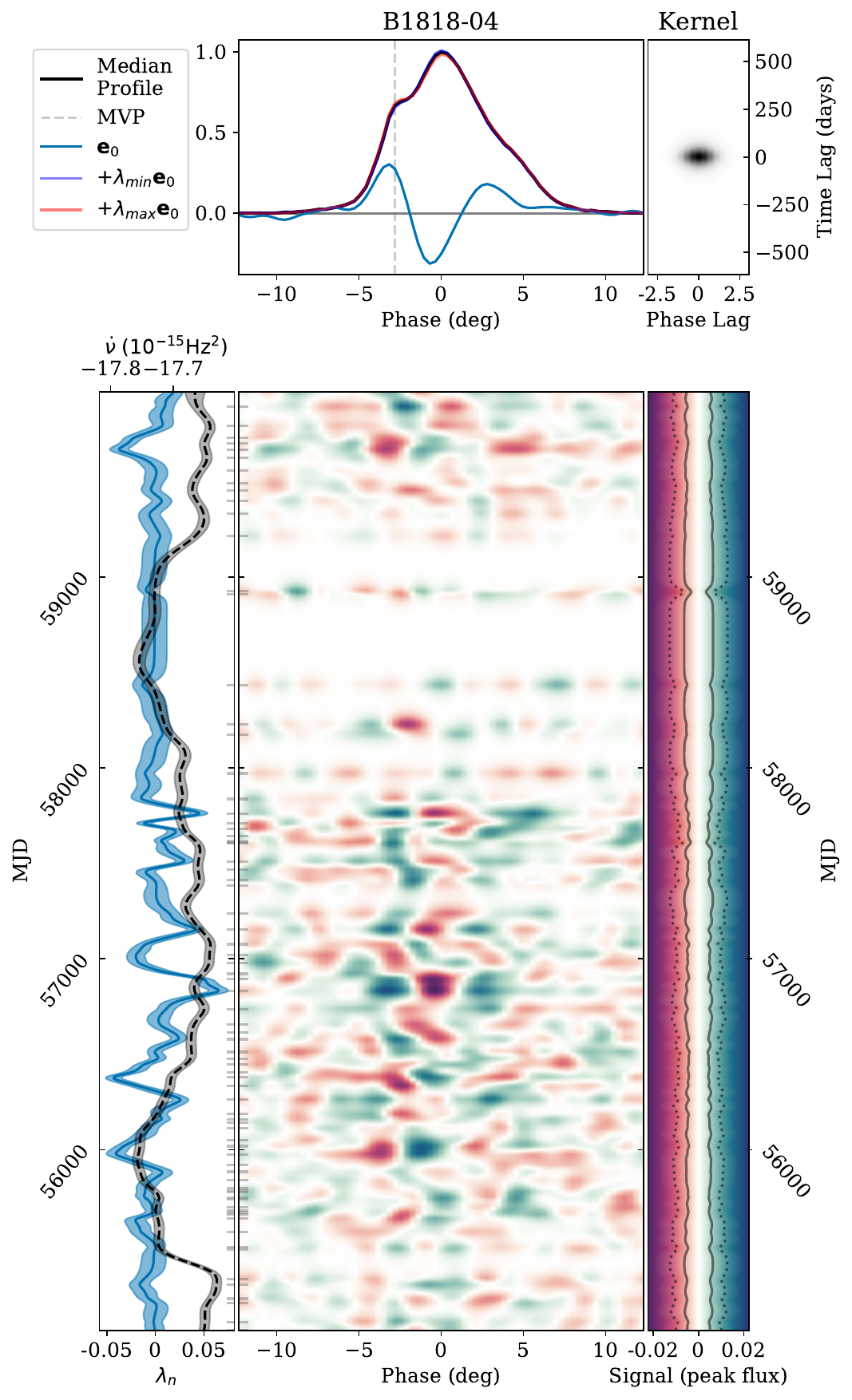}
    \caption{\label{rainbow1818}The output of \textsc{psrcelery} for PSR B1818$-$04. Otherwise as Figure~\ref{rainbow0740}.}
\end{figure}

Figure \ref{rainbow1818} shows the \textsc{psrcelery} output for PSR B1818$-$04.
This pulsar shows fluctuations in \nudot\ on timescales of 7--10 years.
S22 report no evidence for significant profile change, however our analysis appears to show some time-correlated changes in the profile shape with a `M' or `W' shape eigenprofile, and reminiscent of the changes observed in several other pulsars in this sample.
As with PSR B0950+08 the correlation with \nudot\ is not significant so it is not possible to determine the appropriate sign of the shape change.
The observed fluctuations in the pulse shape are again on a shorter timescale than the long-term \nudot\ variations.
It is not clear if there is some underlying relationship between the profile shape variations and the \nudot, which is partly limited because of the low observing cadence on this pulsar.

\subsubsection{PSR B1822$-$09}

\begin{figure}
    \centering
    \includegraphics[width=8.5cm]{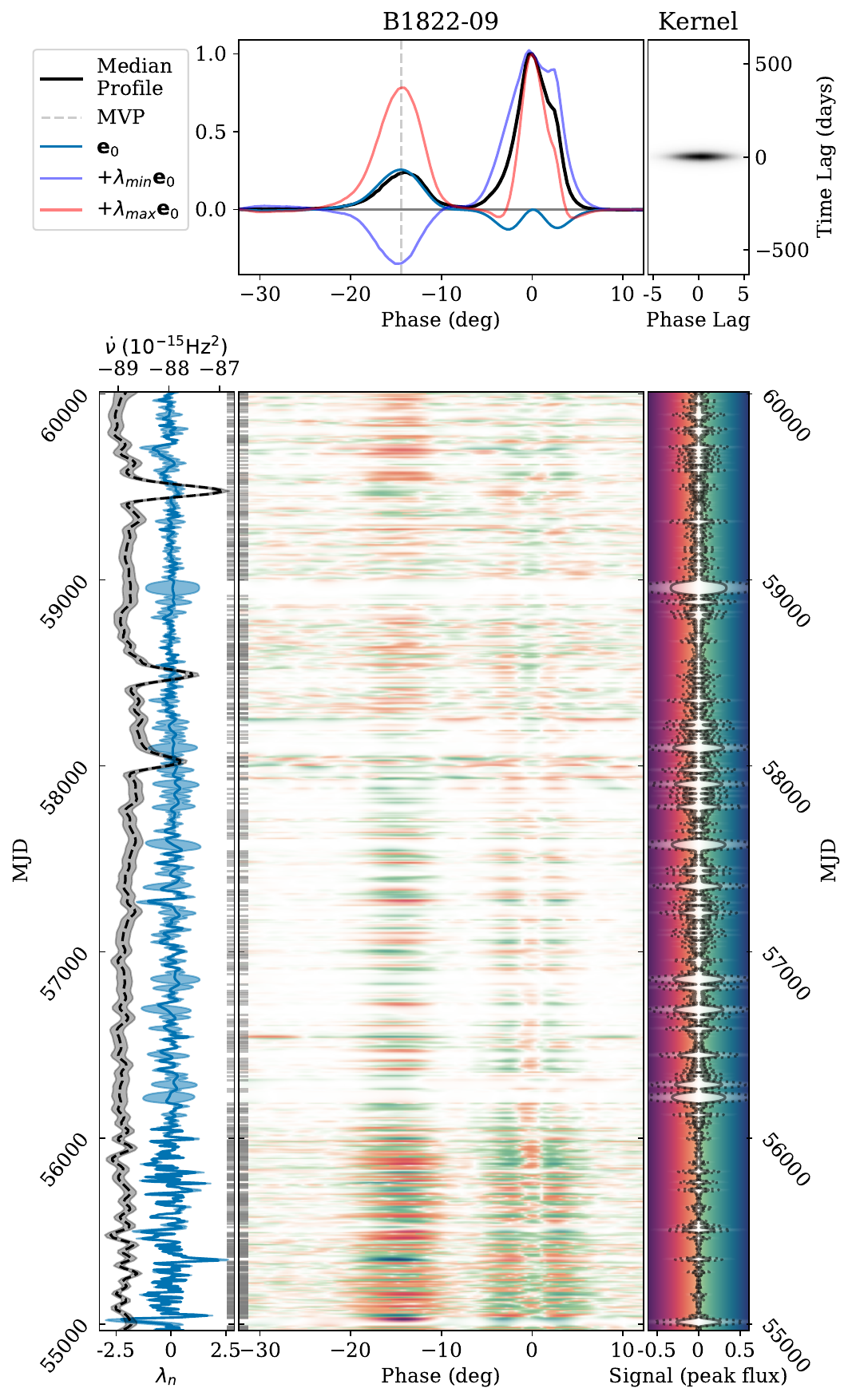}
    \caption{\label{rainbow1822}The output of \textsc{psrcelery} for PSR B1822$-$09 (Main pulse only). Otherwise as Figure~\ref{rainbow0740}.}
\end{figure}
\begin{figure}
    \centering
    \includegraphics[width=8.5cm]{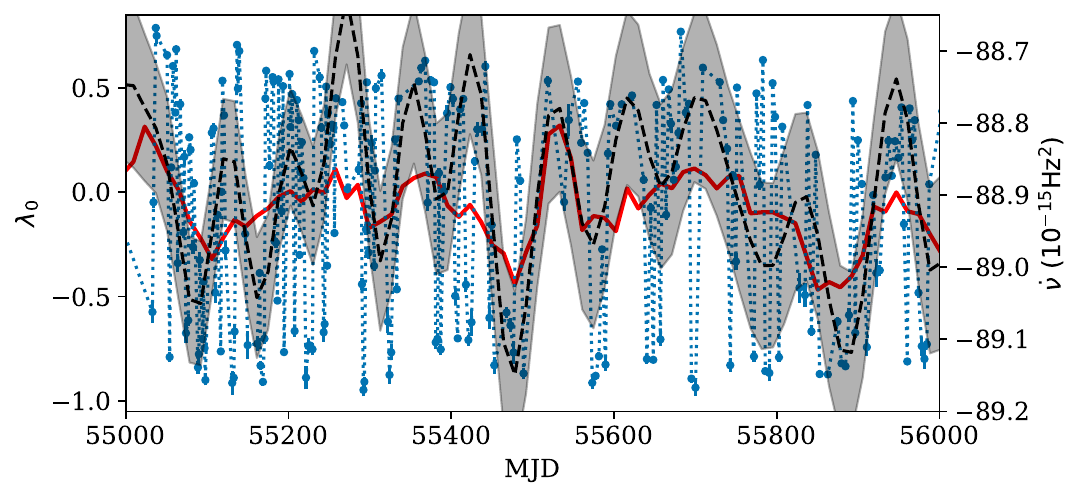}
    \caption{\label{1822zoom} The \nudot\ and PCTS of the first eigenprofile ($\lambda_0$) timeseries for PSR B1822$-$09 between MJD 55000 and MJD 56000. \nudot\ is shown as the black dashed line. The PCTS shown as points is computed directly at each observation epoch to more clearly show the rapid variability and the solid line shows the running mean over 60 day windows.}
\end{figure}

Figure \ref{rainbow1822} shows the \textsc{psrcelery} output for the main pulse of PSR B1822$-$09.
This pulsar is well known to have strong mode changes, where the main pulse switches between a strong single component and a weaker double profile, correlated with the appearance of an interpulse~\citep{fwm81,fw82}.
The mode switching timescale, tens of minutes, is similar to our observation length, and hence the observed pulse shape is the result of a combination of the intrinsic ratio of modes and the chance alignment of the mode changes with our observation epoch.
Hence, the profile shape appears to vary rapidly from observation to observation and the GP timescale reduces towards the minimum bound. 
The \nudot\ of PSR B1822$-$09 is relatively stable for long periods, with sudden short increases, three of which are seen in Figure \ref{rainbow1822}.
During the first two epochs, the profile shape computed by the GP is consistent with the random fluctuations, though the third seems to be coincident with a preference for the two-component mode, which is also the mode associated with more positive \nudot\ in L10/S22.
However, the PCA analysis does not show a particularly strong response during this epoch.
This is partly because the evolution of this pulsar seems to be related more to the intrinsic fraction of time spent in each mode rather than the strength of the shape change, something also seen in PSR B1828$-$11 \citep{stairs19}, but also because the profile variations are dominated by a distinct epoch of more significant fluctuations between MJD 55000 and 56000.

We note that although this coincides with the start of the dataset presented here, inspection of earlier and overlapping data from older instruments suggests that this behaviour is indeed localised to this approximate 1000 day window.
The \nudot\ timeseries also seems to show slightly stronger fluctuations during this period, at a timescale close to the limit which we can measure.

Figure \ref{1822zoom} shows the \nudot\ timeseries and the PCTS computed from the first eigenprofile during this period.
We find only a weak correlation ($r\sim 0.3$) between the PCTS (with or without smoothing) and \nudot, but the figure is certainly suggestive that the small fluctuations in \nudot\ may well be related to changes in the profile shape.
For this pulsar we suspect that the GP modelling of the pulse shape variations may not be the most effective way to identify the important properties, and so whilst further detailed study of this pulsar is ongoing, we find that this is outside the scope of this paper.

\subsubsection{PSR B1826$-$17}
\begin{figure}
    \centering
    \includegraphics[width=8.5cm]{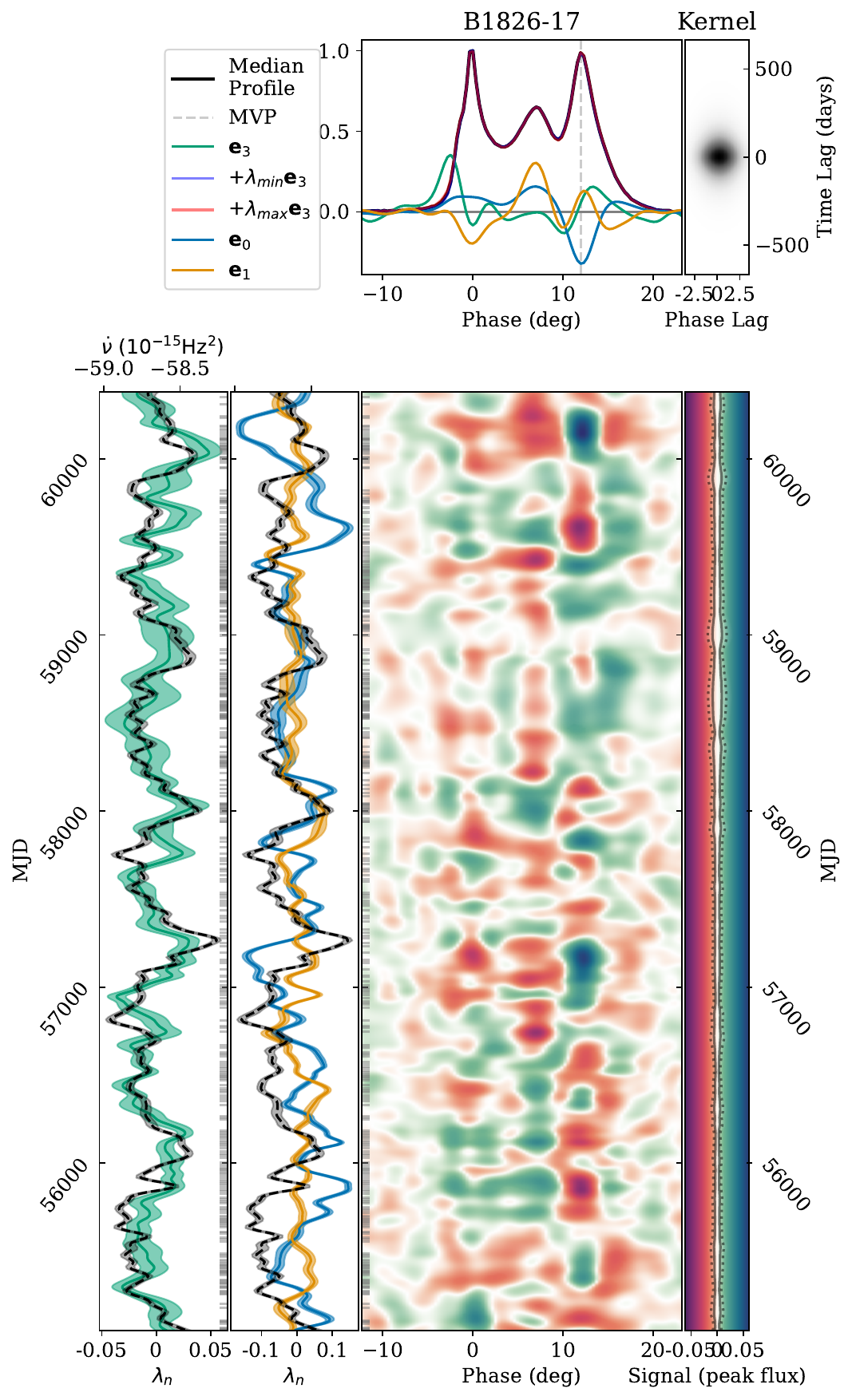}
    \caption{\label{rainbow1826}The output of \textsc{psrcelery} for PSR B1826$-$17. Otherwise as Figure~\ref{rainbow0740}, except an additional leftmost panel showing the PCA results for the fourth eigenprofile.}
\end{figure}
Figure \ref{rainbow1826} shows the \textsc{psrcelery} output for PSR B1826$-$17.
For this pulsar, the most apparent profile variation is a small change in the relative intensity of the peak in the trailing component.
In this case the first eigenprofile, which primarily picks up this change, does not correlate well with the observed \nudot\ timeseries.
However, it is notable that the fourth eigenprofile, which represents only a very subtle change in the leading shoulder and trailing peak, correlates very well with the \nudot\ timeseries.
This suggests that there is an underlying relation between the profile shape and the \nudot\ but that this is masked by larger scale profile shape changes that do not directly correlate with \nudot.
This could also be similar to PSR B0740$-$28, where the relationship between profile shape and \nudot\ is not constant over our observing time, though the boundaries in behaviour are not clear if this is the case.

\subsubsection{PSR B1828$-$11}
\begin{figure}
    \centering
    \includegraphics[width=8.5cm]{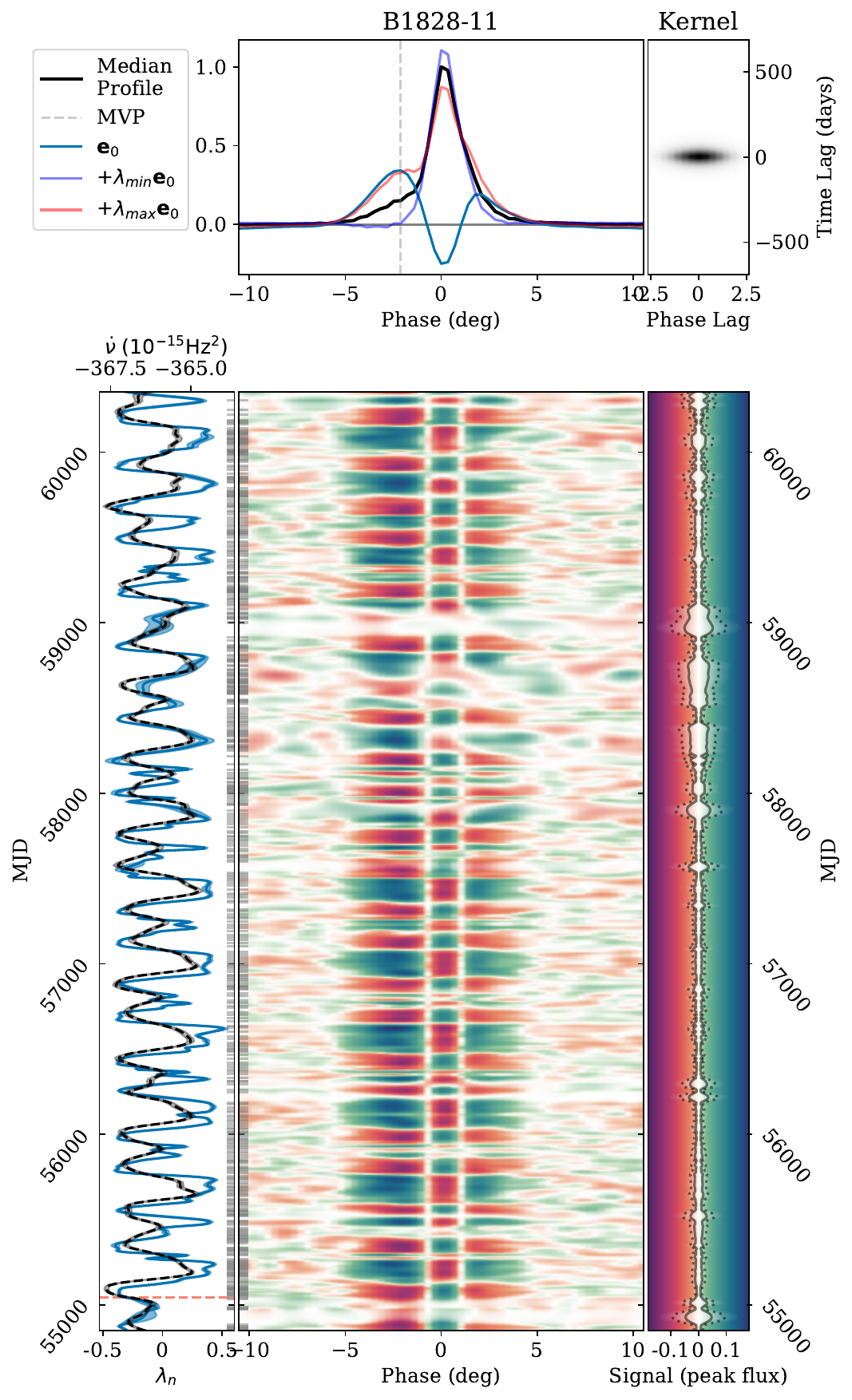}
    \caption{\label{rainbow1828}The output of \textsc{psrcelery} for PSR B1828$-$11. Otherwise as Figure~\ref{rainbow0740}.}
\end{figure}
PSR B1828$-$11 has long been known to show correlation between the pulse shape and the \nudot, both of which are highly periodic on timescales of 500 and 250 days \citep{stairs00}.
Initially this was thought to be evidence for neutron star free precession, but more recent studies suggest that free precession is not required \citep{stairs19}, and L10 suggest this is likely to be the same phenomenon as with the other pulsars in the sample.
Figure \ref{rainbow1828} shows the \textsc{psrcelery} output for PSR B1828$-$11.
As expected, we find a strong correlation between pulse shape and \nudot\ with a substantial change in the ratio of the peak to the shoulders of the profile.
The first eigenprofile correlates with \nudot\ with a correlation coefficient of 0.77.
We note that the periodic oscillations that have been extremely stable to date show a change in character after approximately MJD 58500, where the characteristic double-peaked shape seems to be disrupted for a few cycles.
Nevertheless, the profile shape variations still seem to track the \nudot\ well over this period.

\subsubsection{PSR B2035+36}
\begin{figure}
    \centering
    \includegraphics[width=8.5cm]{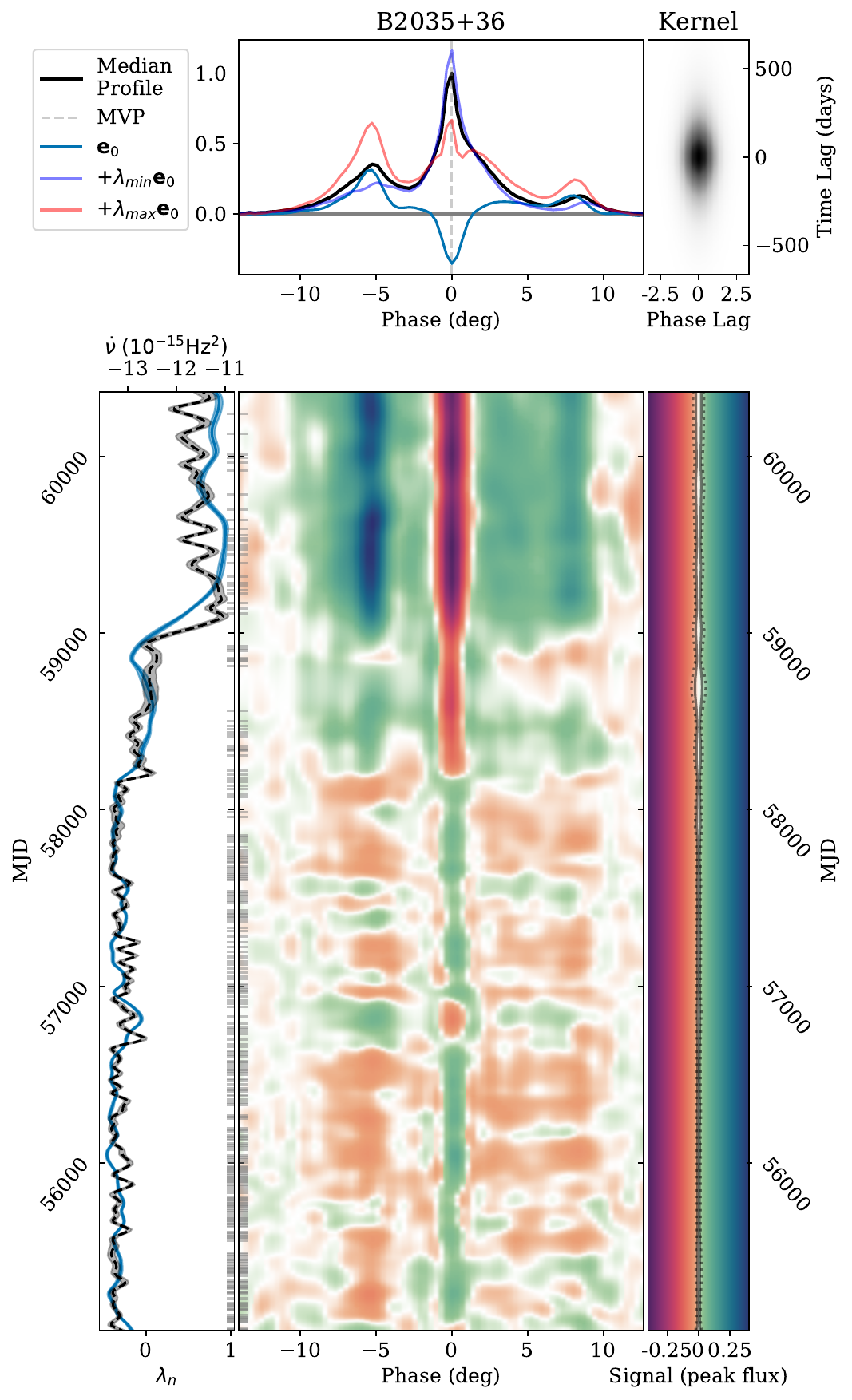}
    \caption{\label{rainbow2035}The output of \textsc{psrcelery} for PSR B2035+36. Otherwise as Figure~\ref{rainbow0740}.}
\end{figure}

PSR B2035+36 showed a large step-change in \nudot\ around MJD 53000, from $-1.17\times 10^{-16}\,\mathrm{Hz^2}$ to $-1.33\times 10^{-16}\,\mathrm{Hz^2}$ (L10, S22).
This was associated with a change in the shape of the profile, characterised by a pronounced decrease in the central component relative to the wings.
Since the publication of S22, this pulsar has now exhibited two step changes of \nudot\ in the opposite sense, i.e. a decrease in spin-down rate.
The first of these steps occurred around MJD 58200, with \nudot\ changing to $-1.25\times 10^{-16}\,\mathrm{Hz^2}$, followed by a larger change to  $-1.15\times 10^{-16}\, \mathrm{Hz^2}$ around MJD 59000, slightly overshooting the originally observed value.
Our analysis (Figure \ref{rainbow2035}) of the profile shape variations shows a clearly correlated ($r=0.92$) change in the profile shape back towards the original shape.
The first eigenprofile is broadly `M' shaped, though the outer components are more widely separated than the other pulsars we have categorised thus far.
The \nudot\ was observed to be in the original state for nearly 20 years before the first transition occurred, and there was a $\sim16$ years period before the return transition.
If these profile changes are part of a quasi-periodic process as seen in some other pulsars, then the timescale for a complete cycle is likely to be longer than 30 years.

\begin{figure}
    \centering
    \includegraphics[width=8.5cm]{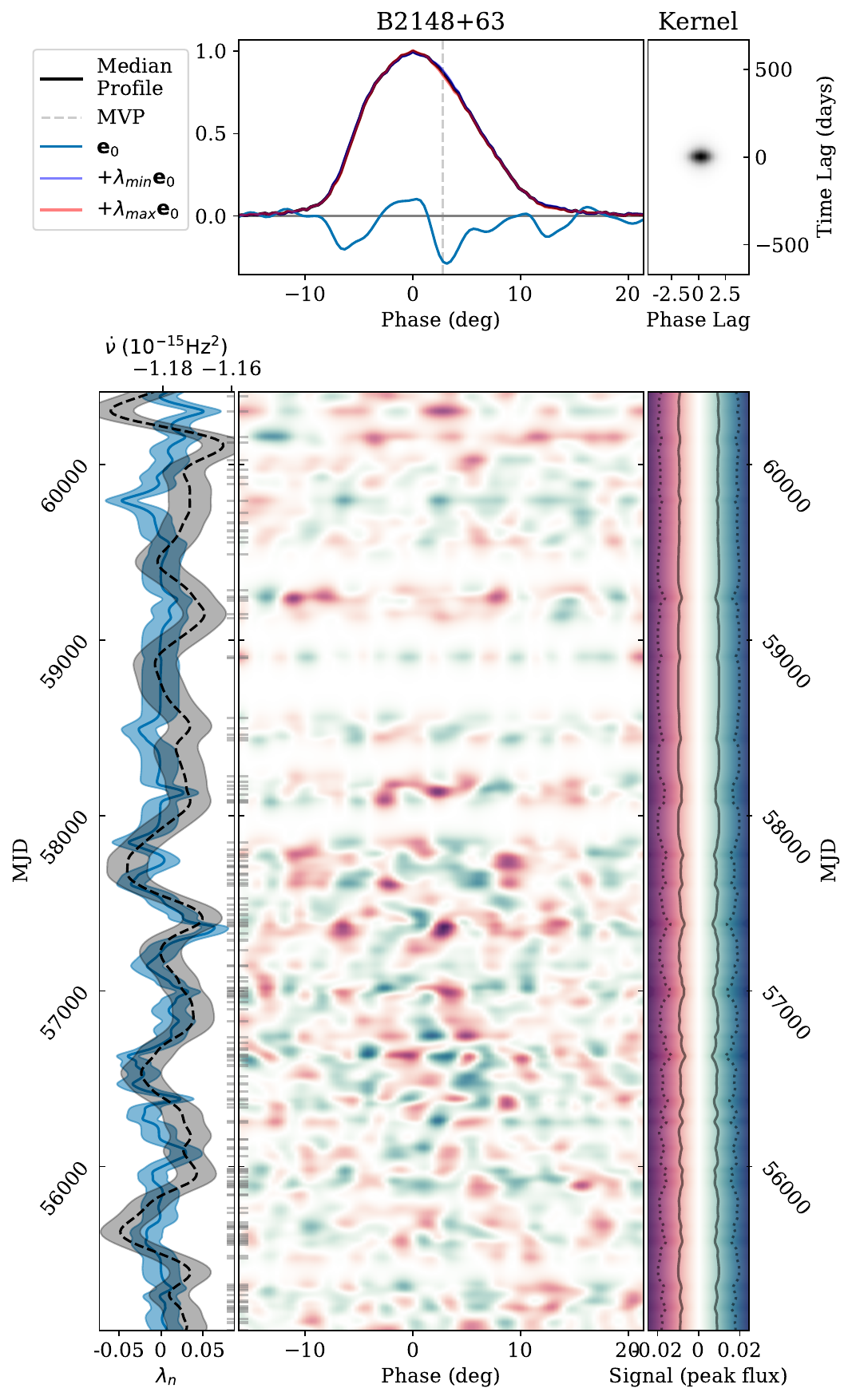}
    \caption{\label{rainbow2148}The output of \textsc{psrcelery} for PSR B2148+63. Otherwise as Figure~\ref{rainbow0740}.}
\end{figure}

\subsubsection{PSR B2148+63}
S22 did not find any profile changes associated with PSR B2148+63, however it has strong quasi-periodic fluctuations in \nudot.
Our analysis (Figure \ref{rainbow2148}) does show barely significant time-correlated fluctuations, though not as clearly structured as in many other pulsars, and the amplitude of changes is only around 1--2\% of the peak intensity.
The GP reconstructed difference map looks quite noisy, however the PCTS does show some correlation with \nudot\ ($r=0.4\pm0.2$), particularly between MJD 56000 and 58000.
There are other fluctuations in \nudot\ that do not seem to be well captured by the PCTS, however we note that the cadence of observations is much poorer in the second half of the dataset and it is possible that here noise from intrinsic variations in individual observations dominates.
The eigenprofile is also hard to interpret and not clearly identifiable as `M' or `W' shape.
Overall, it does seem plausible that there is a correlation between the profile shape and the \nudot, but firm conclusions are hard to draw given the low significance of the profile shape fluctuations.

\subsubsection{PSR J2043+2740}
\begin{figure}    \centering
    \includegraphics[width=8.5cm]{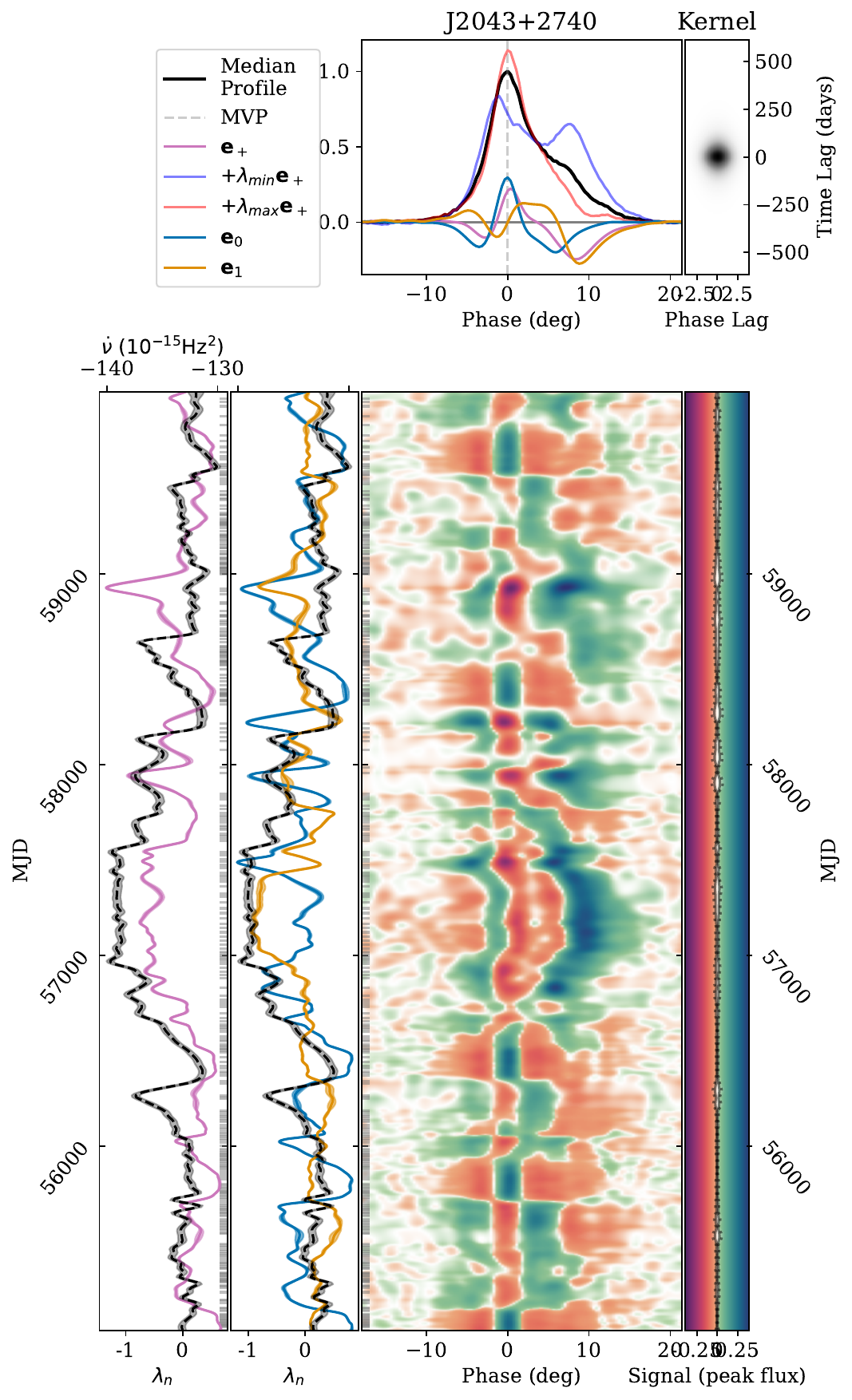}
    \caption{\label{rainbow2043}The output of \textsc{psrcelery} for PSR J2043+2740. Otherwise as Figure~\ref{rainbow0740}, except an additional leftmost panel shows the correlation of the PCTS computed from the sum of the first two eigenprofiles.}
\end{figure}
Both L10 and S22 report long-term profile changes in PSR J2043+2740 correlated with \nudot.
Our analysis (Figure \ref{rainbow2043}) shows similar behaviour and there is clearly some correlation between \nudot\ and the observed profile changes.
However, the first PCTS does not correlate with \nudot\ at all times, particularly around MJD 57000 to 57500.
This is partly because there seems to be more variability in the profile shape, that is not reflected in the \nudot.
More importantly, the overall change in profile seems to include a change in the separation of the two peaks, which is not strictly able to be represented by a linear transformation as is assumed in the PCA.
For small changes in relative phase, a single eigenprofile can give a good approximation, but for more extreme changes multiple eigenprofiles are needed to capture the changes.
Indeed, the second eigenprofile has a stronger correlation with \nudot, but clearly neither track the result perfectly.
Although the eigenprofile analysis is still somewhat effective at capturing the effects here, likely a different shape parameter will perform better, though we leave this for future work.
We note that the correlation can be improved to $r=0.59\pm0.16$ by summing the first two eigenprofiles (as with PSR B0919+06, and shown in the left most panel of Figure \ref{rainbow2043}), though there are still clearly epochs where significant profile changes are not reflected in the \nudot\ and vice versa.

\subsubsection{PSRs B1714$-$34, B1839+09, B1903+07, B1907+00, B1929+20}
As in S22, we do not find any significant profile changes for PSR B1714$-$34, B1839+09, B1903+07, B1907+00, or B1929+20.

\subsection{Beyond the L10 sample}
Here we show results from a sample of pulsars for which time-correlated pulse shape variation has been noted in the JBO dataset. This is not a complete list and a thorough search through the $\sim1000$ pulsars observed at JBO is underway, but outside the scope of this paper.

\subsubsection{PSR B0105+65}
\begin{figure}
\centering
\includegraphics[width=8.5cm]{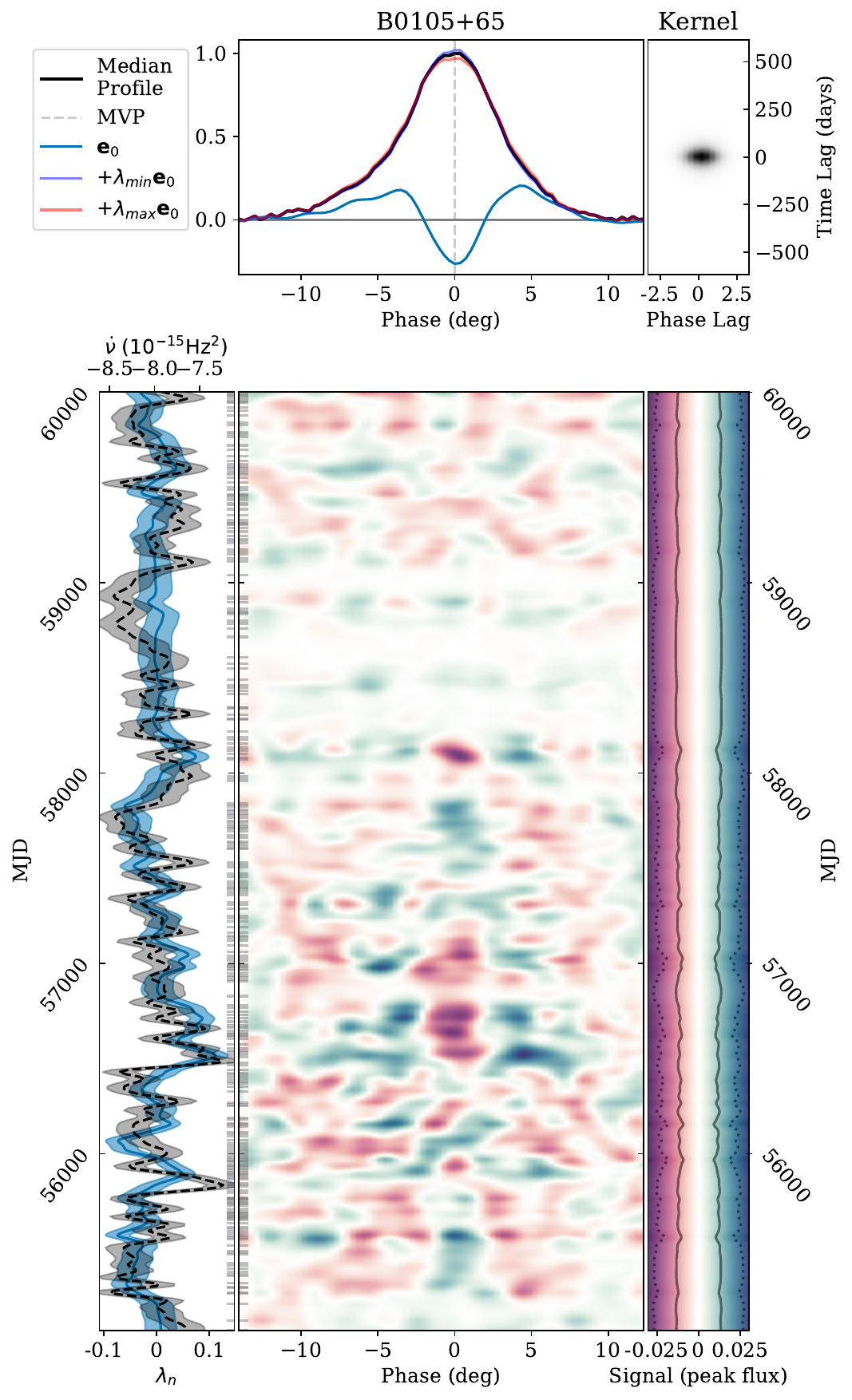}
\caption{\label{rainbow0105}The output of \textsc{psrcelery} for the main pulse of PSR B0105+65. Otherwise as Figure~\ref{rainbow0740}.}
\end{figure}
PSR B0105+65 was noted to show changes in width of $\sim30$\% by \citet{lyneproc}
It has a single component profile with broad tails to the leading and trailing edge.
The \nudot\ shows long-term changes on a $\sim1000$~day timescale as well as more rapid oscillations on a $\sim100$~day timescale.
Our profile analysis (Figure \ref{rainbow0105}) shows some time-correlated profile shape variation, though the GP parameter prefer very short timescales.
The PCTS from the first eigenprofile clearly show similar rapid fluctuations in \nudot\ as well as capturing some of the slower variations.
The correlation coeficient over the full dataset is $r=0.5\pm0.1$, though this seems better at some epochs and the rapid fluctuations make it hard to measure the profile shape when the cadence is low.
The first eigenprofile has a broad and symmetric `M' shape, reminiscent of many of the best pulsars in the L10 sample.
Although the profile typically looks like a single \citep{2022MNRAS.514.3202R}, single-pulse observations \citep{Weltevrede06} show pulse-to-pulse modulation in the outer regions of the profile broadly overlapping with the region of the profile positively correlated with {\nudot}.
Hence, we support the interpretation of this profile as a compact triple.

\subsubsection{PSR B0144+59}
\begin{figure}
\centering
\includegraphics[width=8.5cm]{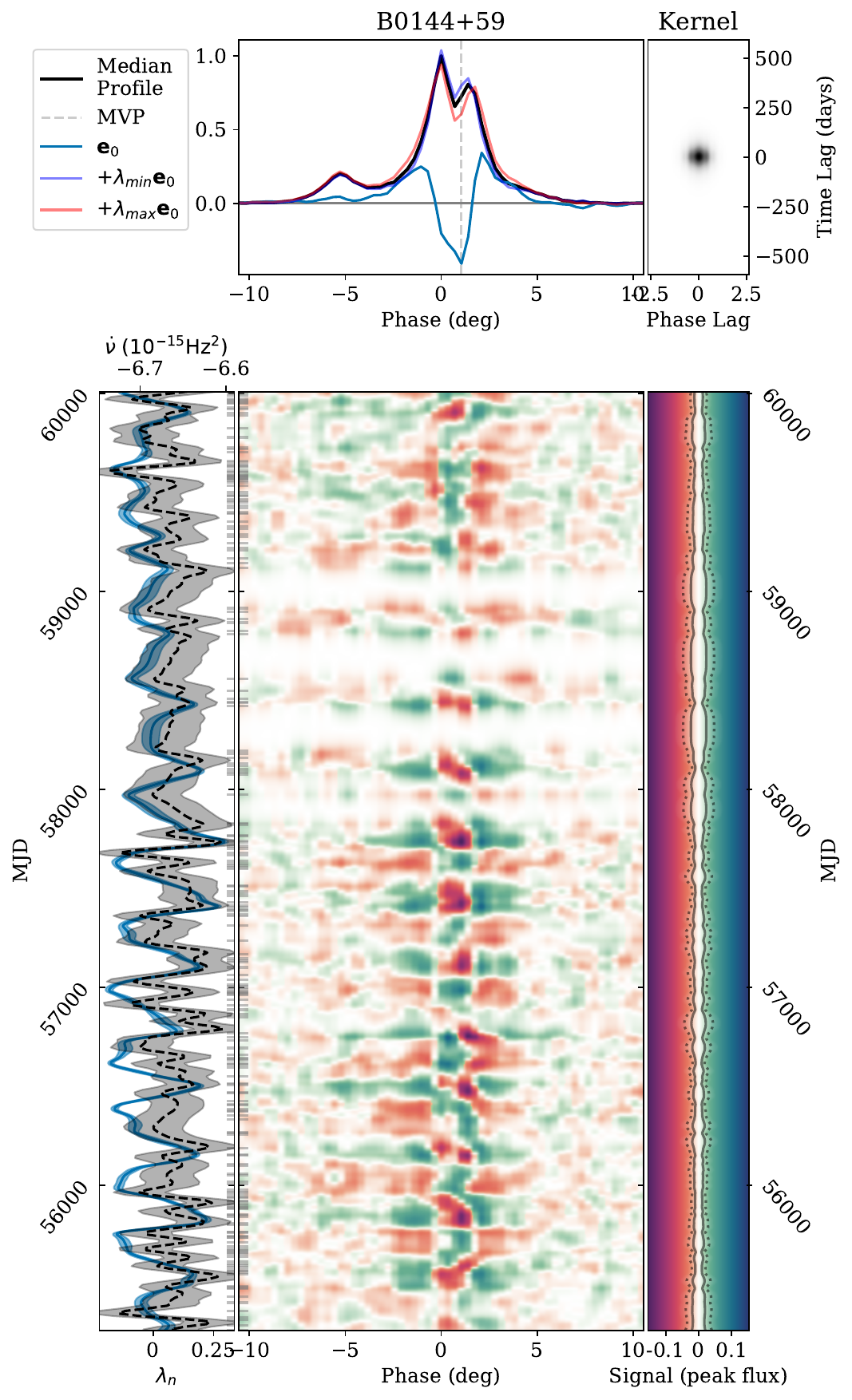}
\caption{\label{rainbow0144}The output of \textsc{psrcelery} for PSR B0144+59. Otherwise as Figure~\ref{rainbow0740}.}
\end{figure}
Highly periodic \nudot\ variations correlated with profile shape changes were reported in this pulsar by \citet{nitu22}.
This pulsar has a double peaked profile with a leading precursor and a smaller trailing component, as well as a much less intense interpulse.
The \textsc{psrcelery} output for the main pulse of PSR B0144+59 (Figure \ref{rainbow0144}) shows the same highly periodic `M'-shaped profile changes in the main pulse as observed in \citet{nitu22}, noting that it is derived from largely the same data. We do not see any variations in the interpulse, likely due to the low signal-to-noise ratio.
The variations seem generally similar to those in L10, and further study, especially at other frequencies may shed more light into the emission structures which are not well understood for this pulsar \citep{2022MNRAS.514.3202R}.
The AFB data for this pulsar have insufficient phase resolution to distinguish the two prominent components in the main pulse.
Comparison of contemporaneous DFB and AFB observations suggests that this is the cause of the discrepancy between the profile published in \citet{gl98} to those in \citet{Weltevrede06} and \citet{sgg+95} noted by \citet{2022MNRAS.514.3202R}, rather than intrinsic mode changes.

\subsubsection{PSR B0611+22}
\begin{figure}
\centering
\includegraphics[width=8.5cm]{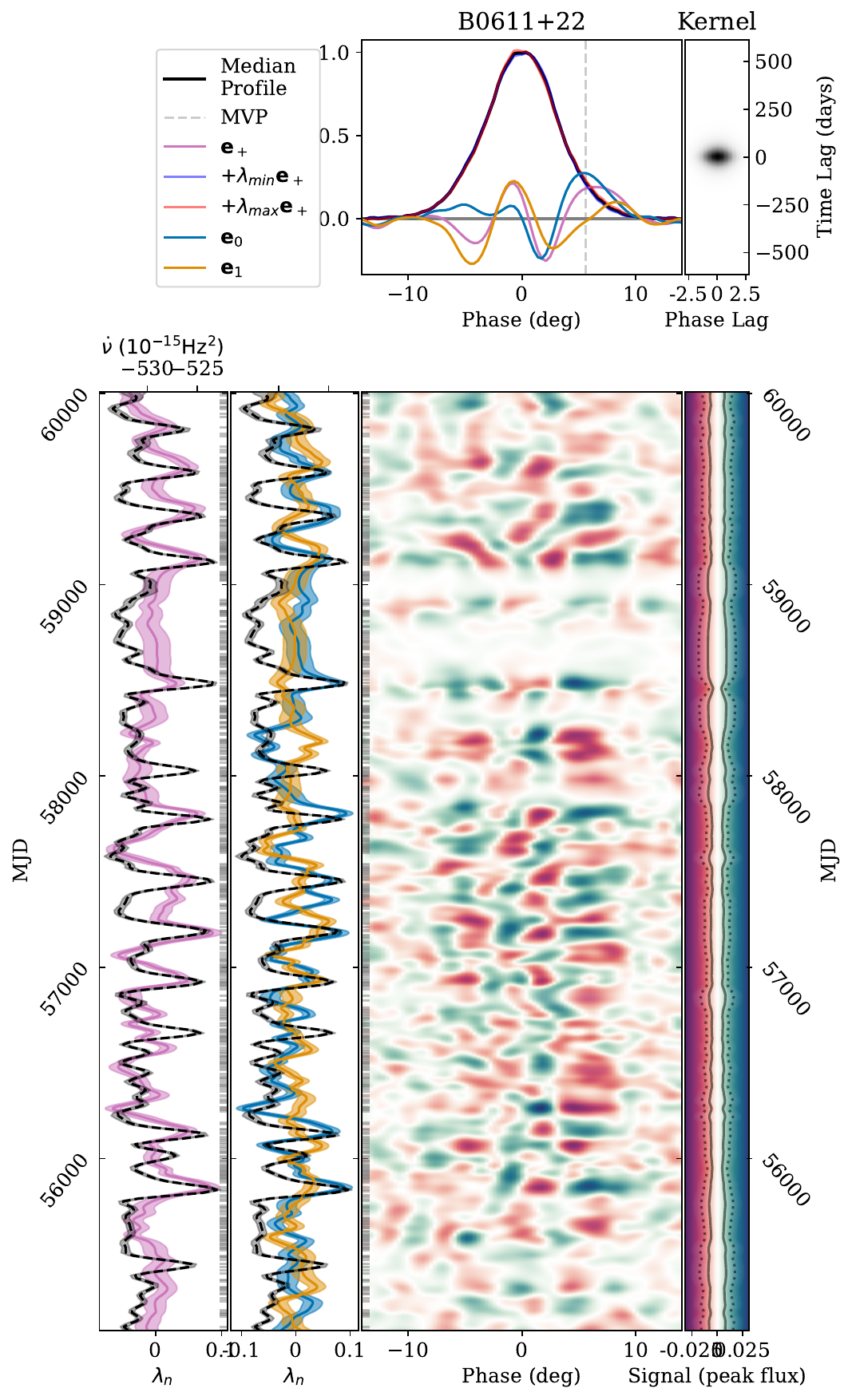}
\caption{\label{rainbow0611}The output of \textsc{psrcelery} for PSR B0611+22. Otherwise as Figure~\ref{rainbow0740}, except an additional leftmost panel shows the correlation of the PCTS computed from the sum of the first two eigenprofiles.}
\end{figure}

PSR B0611+22 has been shown to have quasi-periodic variations in its single-pulse properties on timescales of hours, and showing two distinct emission modes \citep{Seymour14,Rajwade16}.
One mode is characterised by `busrts' of high intensity profiles, which are offset in phase relative to the weak mode.
\citet{sun22} show that the polarised emission also varies slightly between the two modes.  
Figure \ref{rainbow0611} shows the \textsc{psrcelery} output for PSR B0611+22.
The first eigenprofile shows a very subtle change in the trailing edge, similar to PSR B1540$-$06, and consistent with the profile changes being influenced by the presence of the busrting mode.
The PCTS from both the first and second eigenprofiles correlate well with \nudot, both showing short duration switches in state.
As with PSR B0919+06, we find that there is an improvement in the correlation when taking the sum of the first and second eigenprofiles.
The first eigenprofile principally picks up variations in the trailing shoulder of the profile.
The second eigenprofile has a classic `M' shape, with only a small contribution from the shoulder.
Overall, the variations seem consistent with an `M' shape plus an additional rise in the trailing edge of the profile.

\subsubsection{PSR B0626+24}
\begin{figure}
\centering
\includegraphics[width=8.5cm]{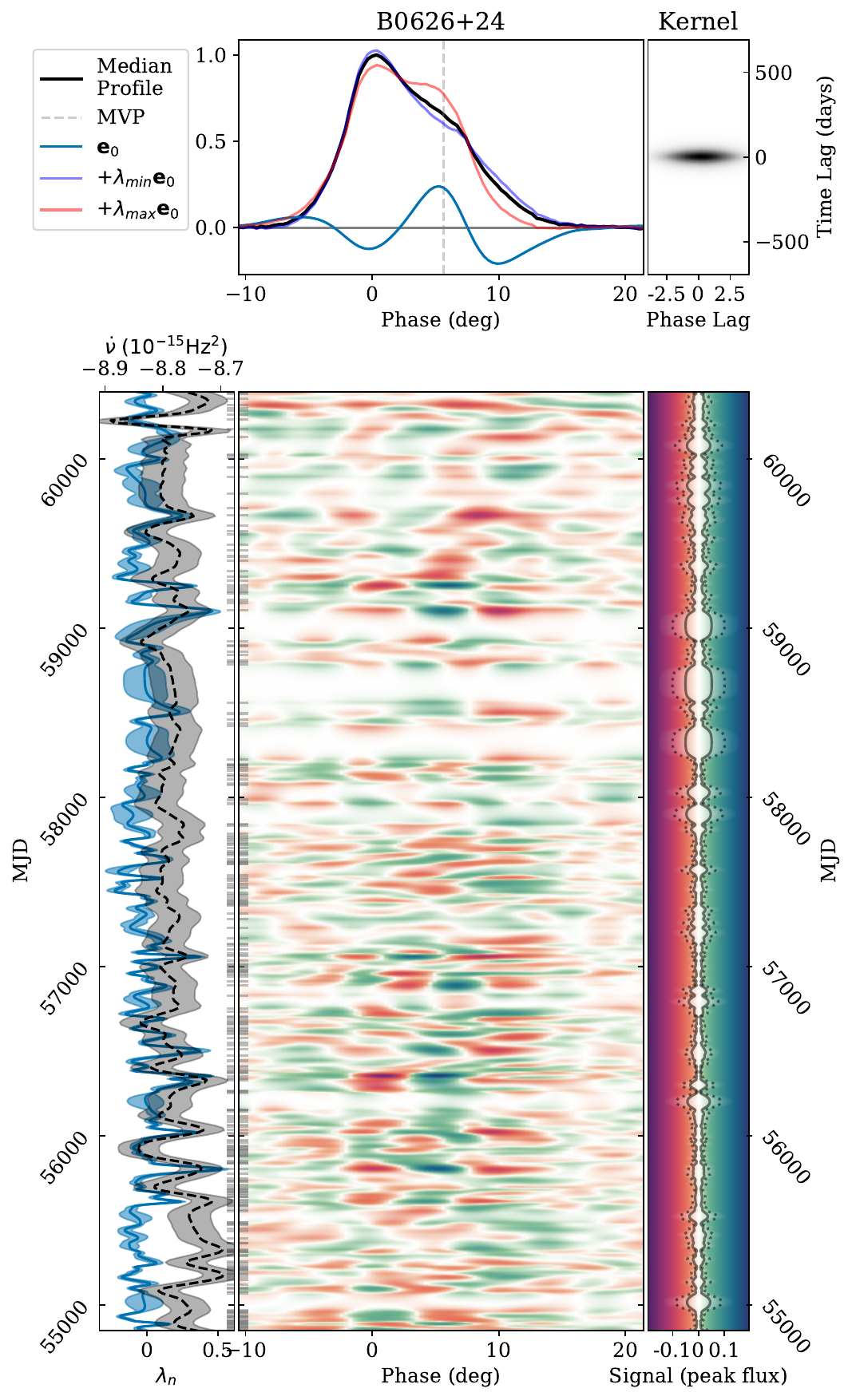}
\caption{\label{rainbow0626}The output of \textsc{psrcelery} for PSR B0626+24. Otherwise as Figure~\ref{rainbow0740}.}
\end{figure}

PSR B0626+24 has not previously been reported to exhibit time correlated profile variations.
Figure \ref{rainbow0626} shows the \textsc{psrcelery} output for PSR B0626+24.
The pulse shape varies rapidly and although the overall correlation with \nudot\ is not very strong ($r=0.13\pm0.14$), there are clearly some transient features that align, particularly around MJD 55700 to 56500.
The eigenprofile that is associated with the larger spin-down rate is `W' shaped, though the very leading edge of the profile is also positive, and hence not so simple to classify.
Studies of this pulsar at other frequencies also suggest a complex overlapping of components and either a core-cone triple structure or two nested emission cones, which are much more symmetric at higher frequencies \citep{Olszanski19}. 
The variations seen here certainly support this picture of complex overlapping components, and observation of the profile shape changes at other frequencies may shed more light on the underlying emission structure.

\subsubsection{PSR B1740$-$03}
\begin{figure}
\centering
\includegraphics[width=8.5cm]{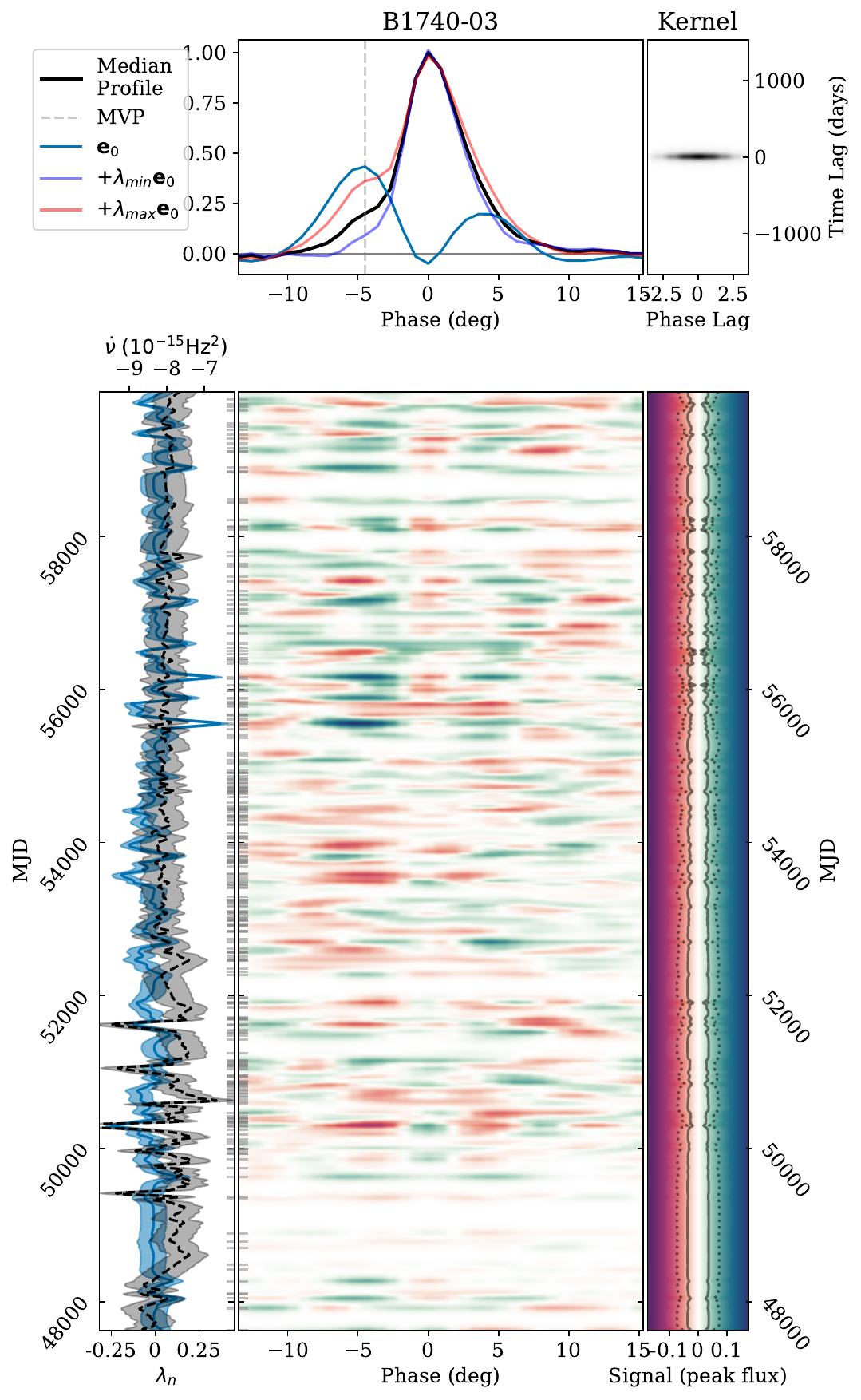}
\caption{\label{rainbow1740}The output of \textsc{psrcelery} for PSR B1740$-$03. This includes both AFB and DFB data. Otherwise as Figure~\ref{rainbow0740}.}
\end{figure}

PSR B1740$-$03 was reported to show variation in the intensity of its components in \citet{lyneproc}, though no profiles are shown.
It exhibits a remarkable change in the spin-down properties between the early data (up to MJD $\sim 53000$) and in more recent data.
In early data, the {\nudot} is characterised by short duration sharp increases in $|\dot{\nu}|$, then seems to transition to a much more stable value.
The {\nudot} is largely stable over the time covered by the DFB data, and hence for this puslar we have analysed a combined dataset of both AFB and DFB data.
To facilitate this, we resample the DFB data to the resolution of the AFB data, and the results are shown in Figure \ref{rainbow1740}.
The transient {\nudot} events in the early data do perhaps have some corresponding changes in profile shape, but overall the correlation between profile shape and {\nudot} is very weak, and many profile shape changes do not seem to have corresponding changes in {\nudot}.

\subsubsection{PSR J1740+1000}
\begin{figure}
\centering
\includegraphics[width=8.5cm]{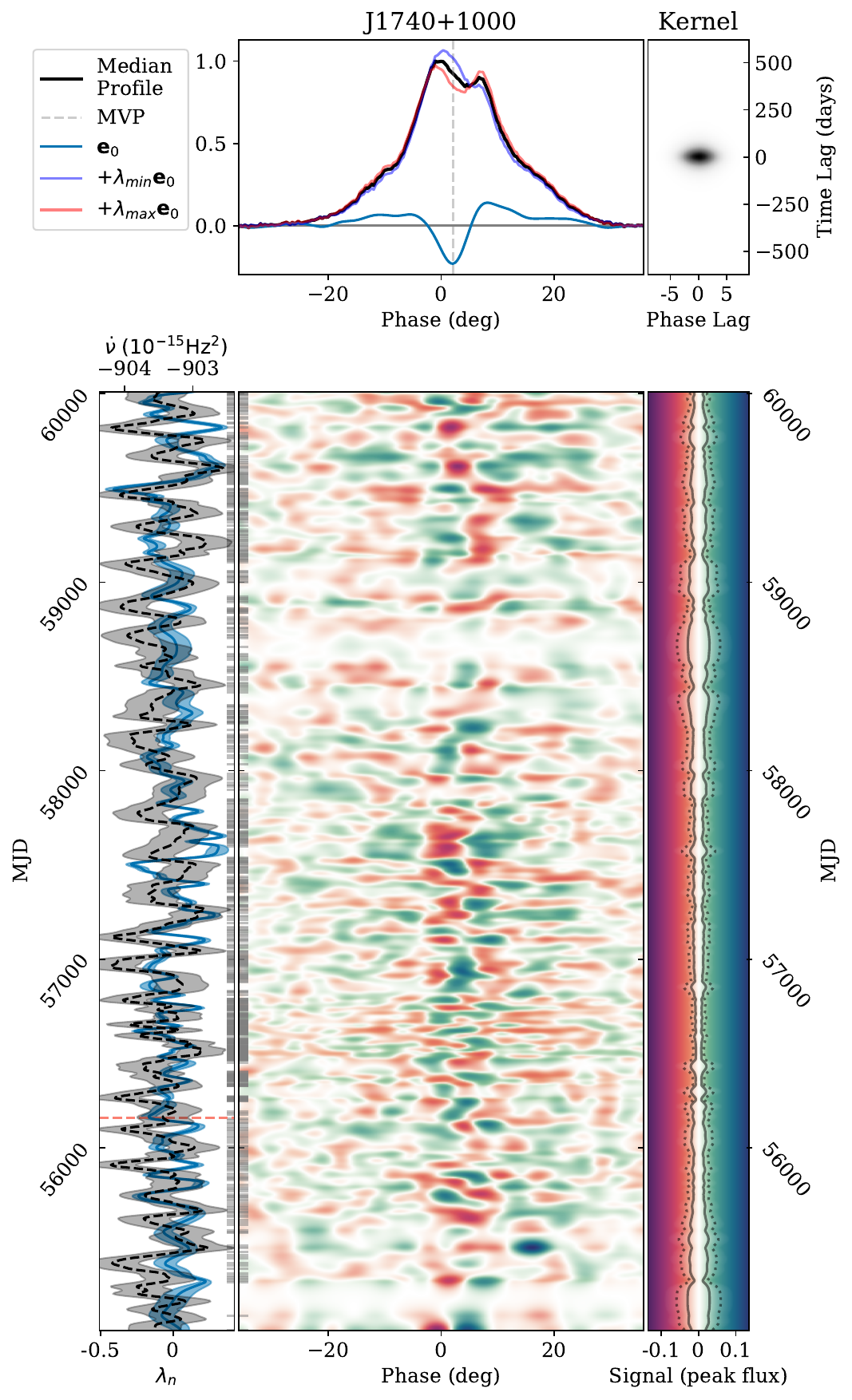}
\caption{\label{rainbowJ1740}The output of \textsc{psrcelery} for PSR J1740+1000. Otherwise as Figure~\ref{rainbow0740}.}
\end{figure}

PSR J1740+1000 is a young, energetic pulsar with X-ray and gamma-ray emission \citep{McLaughlin02,Rigoselli22}.
No previous searches for time-correlated profile shape changes are reported in the literature.
This pulsar shows quasi-periodic \nudot\ variations on a $\sim180$~day timescale.
Our analysis of the profiles (Figure \ref{rainbowJ1740}) also shows significant fluctuations in profile shape on a similar timescale, though not as clearly quasi-periodic.
However, although the two processes seem to have similar characteristics, there is not any significant correlation between the PCTS and {\nudot} ($r=0.08\pm0.13$).
In some ways this pulsar behaves similarly to PSR B0740$-$28, where the overall correlation is low, but there is an impression of both positive and negative correlations at different times, though much less clear in PSR J1740+1000.
It is perhaps also worth noting that other properties of PSR J1740+1000 are also similar to PSR B0740$-$28, with very similar period and period derivative, and hence similar spin-down power and characteristic age, and compact $\sim 100\%$ linearly polarised pulse profiles.
Although superficially appearing as a classic core-cone triple profile, at higher frequencies the profile shows multiple overlapping components not dissimilar to the complex profile of PSR~B0740$-$28 \citep{Olszanski19}.

\subsubsection{PSR B1842+14}
\begin{figure}
\centering
\includegraphics[width=8.5cm]{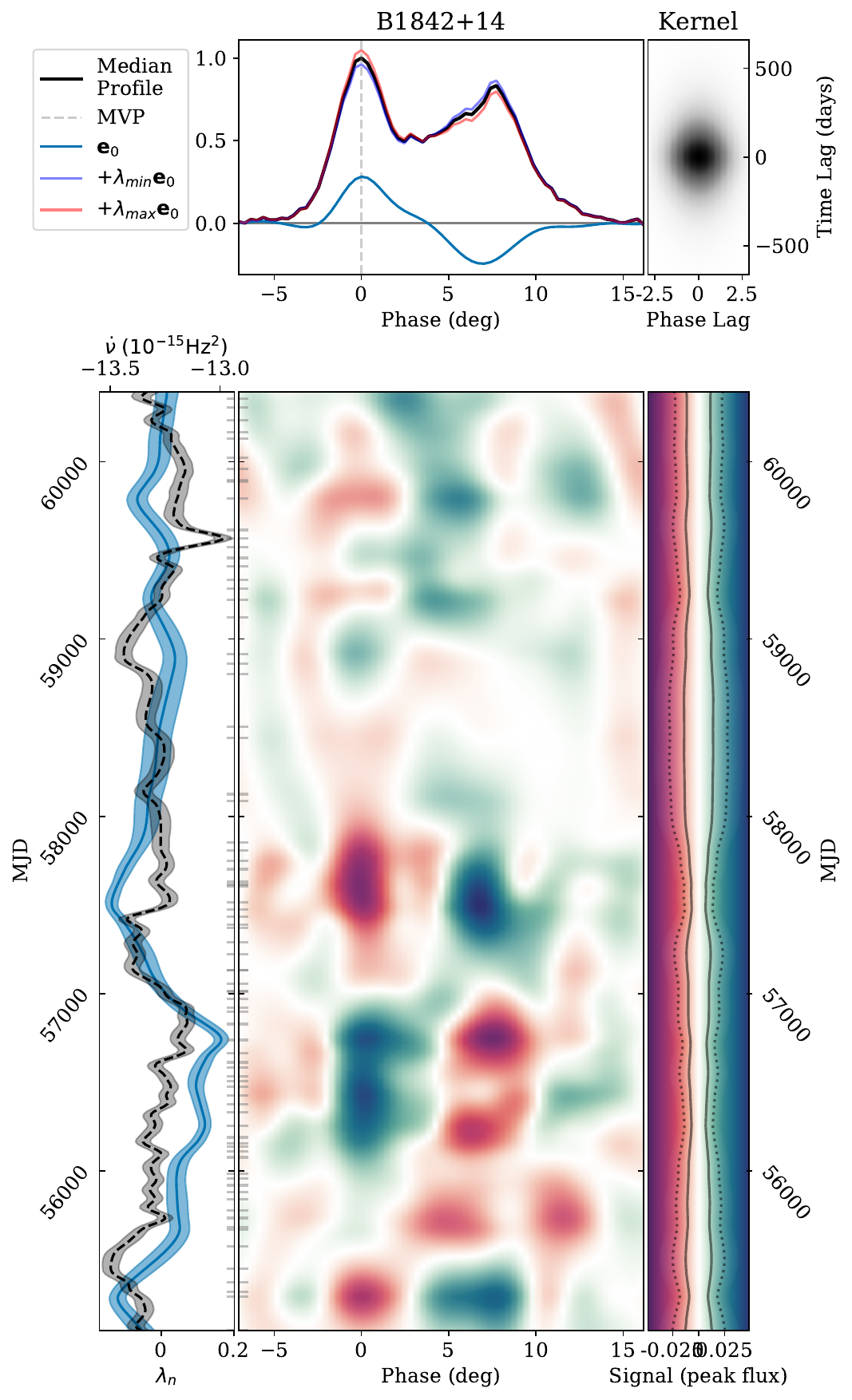}
\caption{\label{rainbow1842}The output of \textsc{psrcelery} for PSR B1842+14. Otherwise as Figure~\ref{rainbow0740}.}
\end{figure}

\citet[hereafter B24]{basu24} report long-term changes in the ratio of the leading and trailing component of PSR B1824+14 in MeerKAT observations.
The same variations can be seen in the JBO DFB data, shown in Figure \ref{rainbow1842}.
There is some hint of correlation between the \nudot\ timeseries and the first eigenprofile, however this isn't supported by a correlation analysis.
B24 find that increasing \nudot\ is associated with a strengthening of the leading component, and we also find this during the period MJD 56000 to 57000, suggesting that the correlation with \nudot\ is not coincidental.
Unlike many of the other examples in our study, PSR B1824+14 shows a change in the relative flux of the leading and trailing components of the profile, however there has been previous suggestions that this is a `partial' cone profile with a missing trailing component. There is some evidence that the observed trailing component may be interpreted as the core component, and hence there is a `missing' trailing component \citep{lm88,mitra11}. 
If this is the case, the eigenprofile would be consistent with an `M' shape.

\subsubsection{PSR B1914+09}
\begin{figure}
\centering
\includegraphics[width=8.5cm]{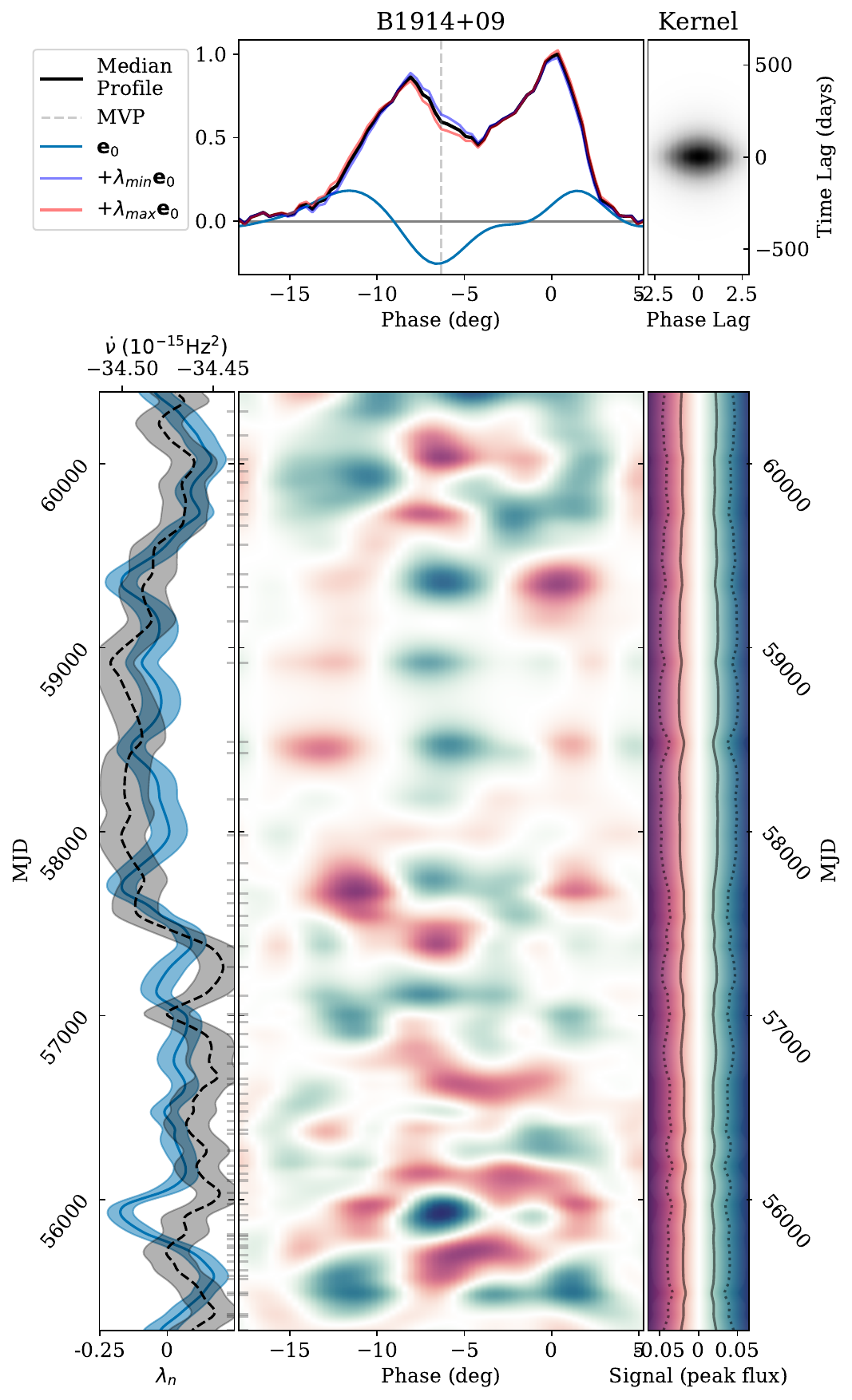}
\caption{\label{rainbow1914}The output of \textsc{psrcelery} for PSR B1914+09. Otherwise as Figure~\ref{rainbow0740}.}
\end{figure}

PSR B1914+09 is also reported in B24, and our analysis can be seen in Figure \ref{rainbow1914}.
We observe the same change in \nudot\ around MJD 59500, as well as an earlier transition around MJD 57800 that precedes the data from B24.
We find that the PCTS correlates well with the \nudot, with the first eigenprofile having a roughly `M' shape, emphasising the edge of the profile over the middle.

\subsubsection{PSR B1917+00}
\begin{figure}
\centering
\includegraphics[width=8.5cm]{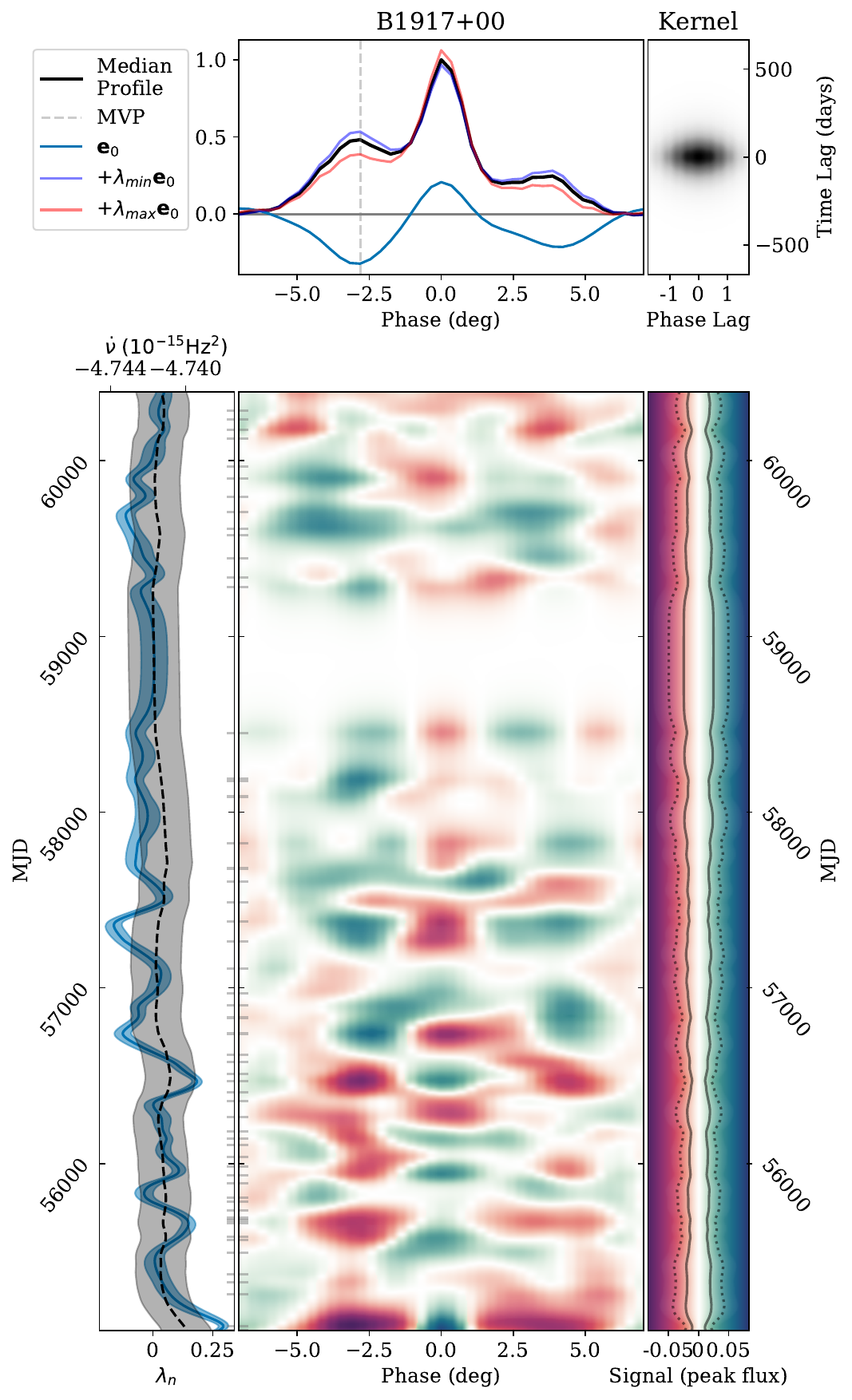}
\caption{\label{rainbow1917}The output of \textsc{psrcelery} for DFB observations of PSR B1917+00. Otherwise as Figure~\ref{rainbow0740}.}
\end{figure}

PSR B1917+00 is also reported to have time-correlated profile changes in B24, and our analysis of this pulsar is presented in Figure \ref{rainbow1917}.
B24 attributes the observed pulse profile changes to the strong pulse-to-pulse variations they observe in the single pulses of this pulsar, and do not find any correlation with \nudot, for which they report no variations over their observing span.
We also find that the \nudot\ does not vary significantly over the timespan of our observations, except at the very start of the dataset, where there is also is a relatively significant profile shape change, though this is only derived from two observations.
We find $r=0.7\pm0.2$, which we find surprisingly strong relative to the impression of the plot.
This is partially due to the two early observations, but even excluding these two gives $r\sim 0.4$, which comes from the subtle long-term variation in the shape and \nudot.
Analysis of the earlier AFB data is more challenging since the phase resolution is poorer ($\sim 10$ phase bins across the width of the pulse).
However, this analysis also shows a weak correlation $r=0.4\pm0.2$ with a similar `W' shaped eigenprofile. Attempting to combine the AFB with resampled DFB data has a  result similar to the AFB data.

\section{Discussion}
\begin{figure}
    \centering
    \includegraphics[width=8.5cm]{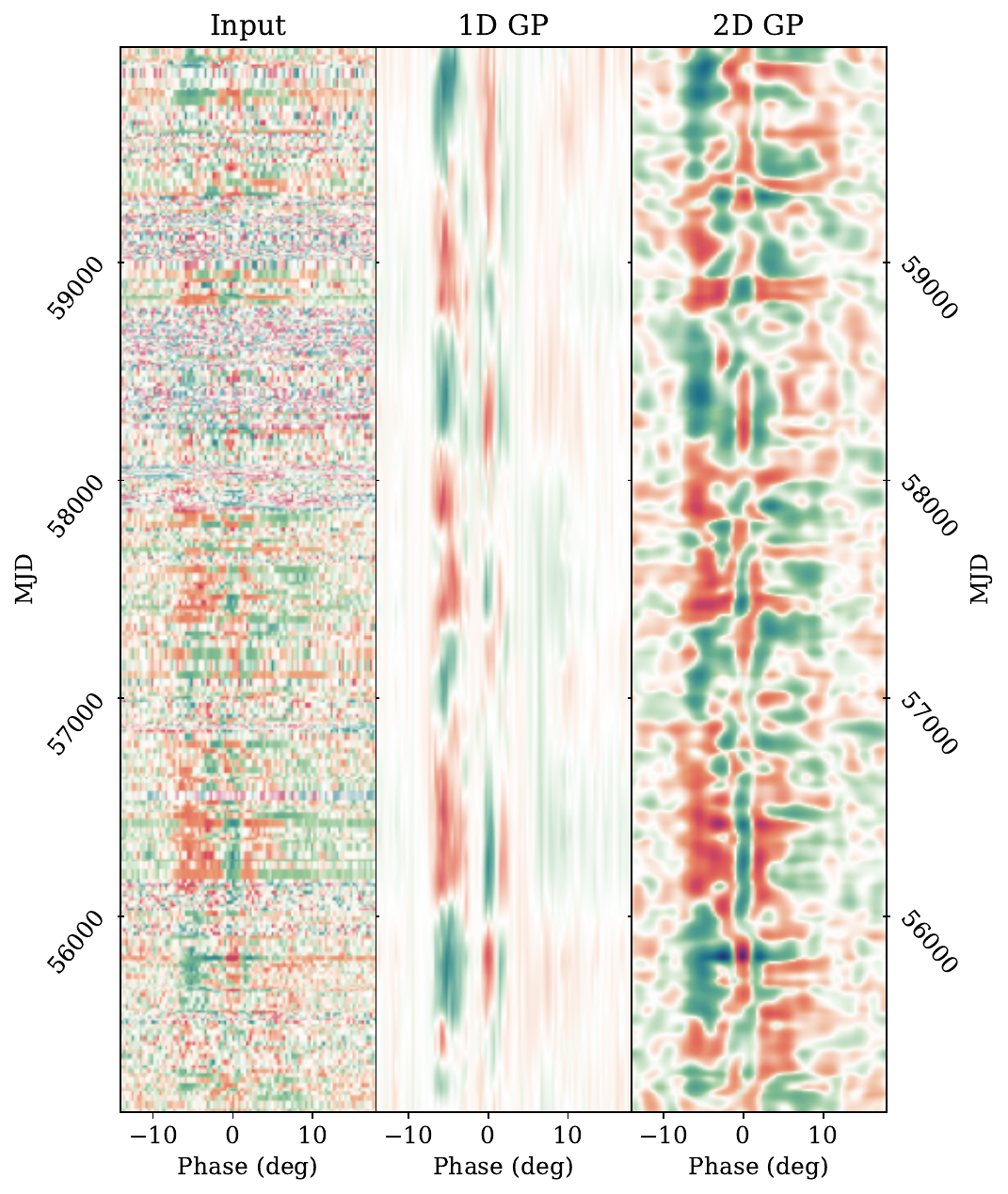}
    \caption{Comparison of the input dataset (interpolated), 1-D GP analysis and 2-D GP analysis for PSR B1642$-$03.}
    \label{compare1642}
\end{figure}

\begin{figure}
    \centering
    \includegraphics[width=8.5cm]{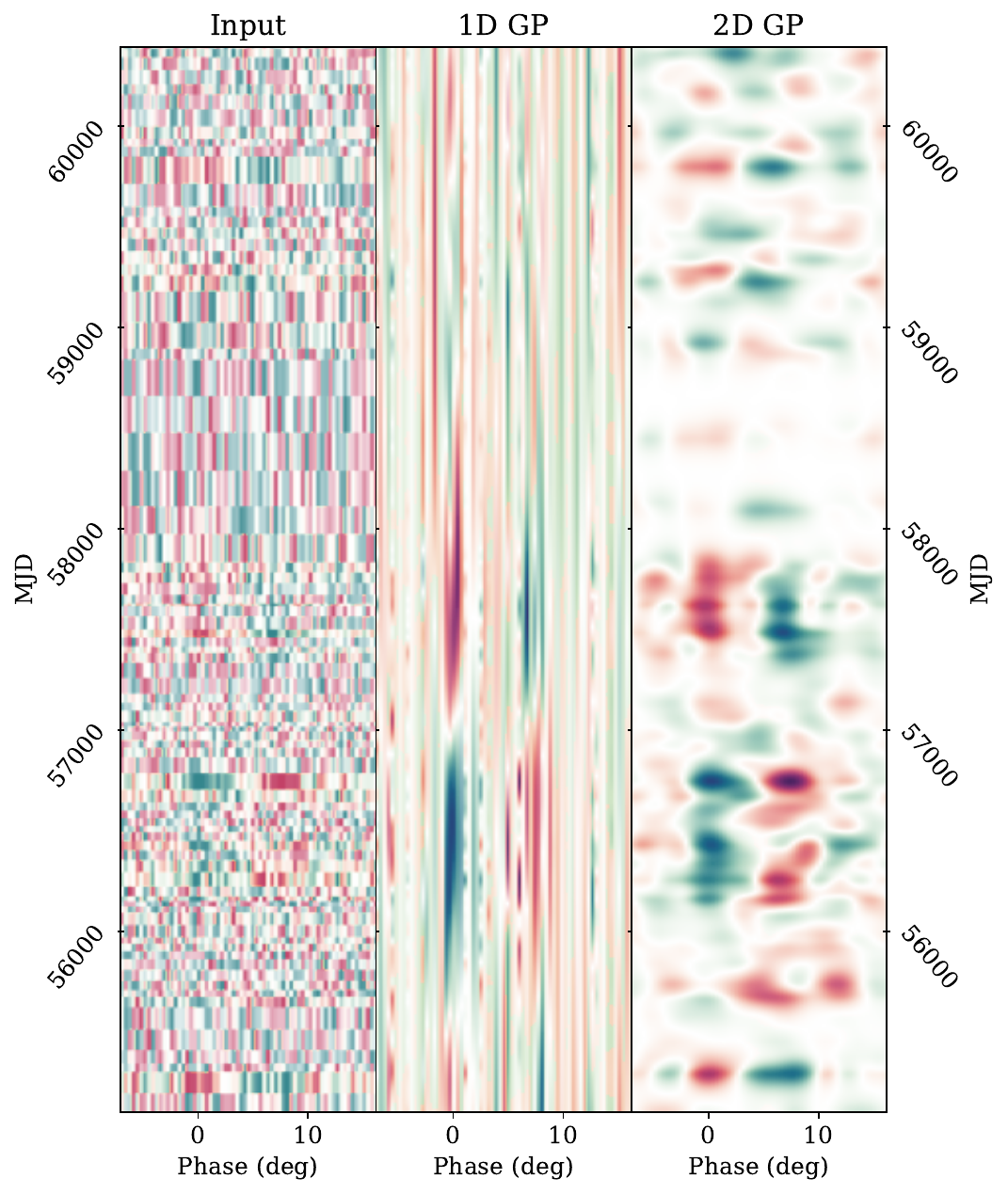}
    \caption{Comparison of the input dataset (interpolated), 1-D GP analysis and 2-D GP analysis for PSR B1842+14.}
    \label{compare1842}
\end{figure}

We have presented the output of \textsc{psrcelery} on 21 pulsars that demonstrate pulse shape variations, finding evidence for a correlation between pulse shape and \nudot\ in 18 cases. This includes 11 pulsars with previously reported correlations, two pulsars (PSR B1826$-$17 and B2148+63) where no shape variations were observed in the L10 or S22 analyses, and five pulsars that had not been previously studied for shape variations in the literature.
In order to make more direct comparisons between the 1-D and 2-D GP approaches, we have processed the same dataset using \textsc{psrcelery} and a 1-D GP model similar to that used by S22. Figure \ref{compare1642} shows the comparison for PSR B1642$-$03, and here both 1 and 2-D GP models clearly identify profile shape changes at the same epochs and phases. Both largely capture the behaviour in the input data, although the `character' of the observed changes is quite different, with some phases seemingly showing little variation in the 1-D case (similar to the result of S22), whereas the 2-D model shows variation across a much wider range of pulse phase. Figure \ref{compare1842} shows the same plot but for PSR~B1842+14. Here the phase-correlated features in the 2-D model gives a clearer impression of the variations, particularly during the later epochs where little coherent structure can be seen in the 1-D model.


In many cases the PCA analysis automatically leads to a good correlation between \nudot\ and the PCTS of the first eigenprofile.
However, this does not work for all pulsars.
For instance, PSR B1826$-$17 clearly has significant profile shape variations, and there is  some correlation with \nudot, but it is not until the fourth eigenprofile that a clear correlation with \nudot\ emerges.
In the cases of B0611+22, B0919+06 and J2043+2740 there is some correlation with the first eigenprofile, but the correlation can be improved by summing the first two eigenprofiles.
It is perhaps not unexpected that the PCA is imperfect for identifying the best shape parameter given that there is no particular expectation that the variations should arise from a simple linear change in profile shape.
Nevertheless it is clear that the PCA analysis is a valuable tool for quantifying the profile variations, especially when applied to the de-noised profile stacks produced from the GP regression model.

It is notable that many of the eigenprofiles have similar shapes, typically increasing \nudot\ (decreasing $|\dot\nu|$) is associated with an increase in the leading and trailing `shoulders' of the profile relative to the peak.
This could be interpreted as an overall widening of the profile, but as also noted in L10, in many cases inspection of the profiles suggests it is more likely changes in the relative intensity of the central component relative to components on the shoulders.
Clearly this is strongly suggestive of a change in the relative strength of core and conal emission, though it is not always easy to identify such structures in pulsars with compact and complex emission profiles.
For the 21 pulsars in Table \ref{results_table}, all lie above or close to the $\dot{E}=10^{32.5}\,\mathrm{erg\,s}^{-1}$ threshold where \citet{corecone_edot} find that the pulsars show strong core emission, and certainly within the typical region of profiles with both cores and cones.
Further investigation of this relationship is certainly warranted, but falls outside the scope of this paper.

Interestingly, whilst 12 of the pulsars show an`M' shaped profile change (i.e. excess core emission), this includes almost all the pulsars with the most obvious correlations with \nudot.
Of the remainder, most show the `W'-shaped inversion of this pattern, with the leading and trailing edges reducing relative to the centre, or equivalently a narrowing of the profile, and two that show more complex shapes.
It is hard to understand why the overall appearance of the correlation should relate to the form of the profile shape changes in this way.
Particularly interesting is PSR B0740$-$28, which seems to change the form of the profile variations on timescales of 1--3 years, including complete inversions between `M' and `W' shapes.
This pulsar has a complex series of nested profile components and it seems likely that the behaviour here is more complex than the change in the relative intensity of core and conal components.

Whilst the application of the 2-D GP modelling followed by PCA analysis is effective at identifying {\nudot}-correlated profile changes, the cases of PSR B0950+08 and B1822$-$09 demonstrates that, especially when the profile changes on timescales of the observing duration, the {\nudot} state is only correlated with time-averaged parameters. It is likely that, as with e.g. PSR B1828$-$11 (L10; \citealp{stairs19}), these pulsars are switching between states faster than we can estimate {\nudot} and it is the fraction of time spent in various states that correlates with the observed {\nudot}.

\section{Conclusions}
We have demonstrated a 2-D GP model for pulsar profile variations as applied to a selection of pulsars from the Jodrell Bank observing programme. This utilises the intrinsic correlation in pulse phase of profile shapes to better model the profile variations. This provides an new tool for visualisation of time and phase correlated profile shape variations, and complements existing 1-D GP methods. We also demonstrate that applying principle component analysis to the GP model can be effectively used to `automatically' characterise the most significant profile shape changes, and that this often correlates with \nudot.

The most prominent profile variations, as determined by PCA, seem to have a characteristic shape that is suggestive of a change in the ratio of `core' and `cone' emission.
A more complete study of the underlying processes behind the profile variations would benefit from full-polarisation and frequency-resolved studies over a wide enough band to identify both the time and frequency evolution of the profile.
Such observations are already ongoing at MeerKAT \citep{basu24,TPA_DR1} and Murriyang \citep{lower25}.
Finally, we note that the pulsars shown here do not represent the full sample of pulsars with profile variations at Jodrell Bank, and hence cannot be used for a population study.
A full exploration of the $\sim1000$ pulsars observed at JBO is the subject of future publications.

\section*{Acknowledgements}
Pulsar research at Jodrell Bank is supported by a consolidated grant (ST/T000414/1 and ST/X001229/1) from the UK Science and Technology Facilities Council (STFC). D.A. acknowledges support from an UKRI fellowship (EP/T017325/1)

\section*{Data Availability}
The data used in this work is available from \url{doi.org/10.5281/zenodo.14005816}, and the analysis can be repeated using the \textsc{psrcelery} software as published at \url{doi.org/10.5281/zenodo.14005776}. The psrcelery software repository is also available at \url{https://github.com/SixByNine/psrcelery}.




\bibliographystyle{mnras}
\bibliography{refs} 








\bsp	
\label{lastpage}
\end{document}